\documentclass[amsmath,amssymb,superscriptaddress,showpacs,twocolumn]{revtex4}
\usepackage{graphicx}
\usepackage{textcomp}
\usepackage{dcolumn}
\usepackage{amsmath}
\usepackage[utf8]{inputenc}
\usepackage{amssymb}
\usepackage{natbib}
\usepackage{booktabs}
\usepackage{multirow}
\usepackage{tabularx}   
\usepackage{tabularray}
\usepackage{gensymb}
\usepackage{lipsum}
\usepackage{blindtext}
\usepackage{mathtools}
\usepackage{tensor}
\usepackage[colorlinks=true, citecolor=blue, urlcolor = blue, linkcolor= red, bookmarks=true]{hyperref}
\usepackage{float}
\usepackage{amsfonts}
\usepackage{mathrsfs}
\usepackage{afterpage}
\usepackage{array}  
\usepackage{mathrsfs}

\begin{document}
\title{Analytical treatment of strong gravitational lensing by rotating black-bounce spacetimes with and without horizons}
\author{Shafqat Ul Islam} \email{shafphy@gmail.com}
\affiliation{Institute for Theoretical Physics and Cosmology, Zhejiang University of Technology, Hangzhou 310023, China}

\author{Qiang Wu} \email{Corresponding author: wuq@zjut.edu.cn} 
\affiliation{Institute for Theoretical Physics and Cosmology, Zhejiang University of Technology, Hangzhou 310023, China}

\author{Tao Zhu} \email{Corresponding author: zhut05@zjut.edu.cn} 
\affiliation{Institute for Theoretical Physics and Cosmology, Zhejiang University of Technology, Hangzhou 310023, China}

\author{Sushant~G.~Ghosh }\email{sghosh2@jmi.ac.in}
\affiliation{Centre for Theoretical Physics, 
	Jamia Millia Islamia, New Delhi 110025, India}
\affiliation{Astrophysics and Cosmology Research Unit, 
	School of Mathematics, Statistics and Computer Science, 
	University of KwaZulu-Natal, Private Bag 54001, Durban 4000, South Africa}
    
\begin{abstract}

Strong gravitational lensing is one of our most direct probes of gravity in the strong-field regime, and horizonless ultracompact objects are expected to leave lensing signatures that are qualitatively different from those of black holes. Recent EHT observations of supermassive compact objects have opened new avenues for testing gravity in the strong-field regime. These observations do not rule out the possibility that some compact objects may be horizonless rather than black holes and may produce distinct gravitational lensing signatures. In this work, we study strong gravitational lensing by rotating black-bounce spacetimes, regular geometries that smoothly interpolate between black holes and traversable wormholes. We analyse two models that extend the Simpson-Visser spacetime, each with a distinct mass function that affects the near-throat structure. We find that horizonless spacetimes can support both an outer unstable photon sphere and an inner stable photon sphere. Images associated with trajectories passing inside the outer photon sphere show larger angular separations and magnifications than those formed outside, in some cases by several orders of magnitude, which may be relevant for future high-resolution observations such as ng-EHT. When we compare with EHT observations of Sgr A$^*$, both the black hole and the horizonless compact object remain consistent when the spin is low. At the same time, differences become more apparent at higher spins within the parameter space considered. We also prove that black-bounce spacetimes can admit closed photon ring structures, in contrast to Kerr naked singularities, which produce an open arc. Altogether, this work is the first analytic study of strong gravitational lensing in rotating horizonless black-bounce spacetimes and provides a framework applicable to other similar compact objects.

\end{abstract}

\maketitle

\section{Introduction}
Strong gravitational lensing is one of the most powerful ways to study gravity in the strong-field regime. It provides direct information about the properties of compact objects and the geometry of spacetime near them \cite{Perlick:2010zh, Virbhadra:1999nm}. Recent observational advances, such as the horizon-scale images of supermassive black holes obtained by the Event Horizon Telescope (EHT) \cite{EventHorizonTelescope:2019dse, EventHorizonTelescope:2022wkp} and the detection of gravitational waves from compact binary mergers \cite{LIGOScientific:2016lio}, have provided powerful new tests of general relativity (GR) and alternative theories of gravity. These observations have confirmed many predictions of Einstein's theory while also motivating the study of exotic compact objects that can mimic black holes and remain compatible with current observations \cite{Cardoso:2016rao, Cunha:2018gql, Carballo-Rubio:2018jzw, Konoplya:2018yrp}.

Despite the significant experimental and observational successes of GR, several theoretical and cosmological challenges indicate that it may not offer a complete description of gravity. The existence of spacetime singularities and the absence of a consistent quantum theory of gravity highlight potential limitations of Einstein’s framework \cite{Ashtekar:2004eh, Kiefer:2013jqa}. On cosmological scales, the standard GR-based cosmological model depends on the introduction of dark matter and a cosmological constant to account for galactic dynamics and the observed accelerated expansion \cite{SupernovaSearchTeam:1998fmf}. These challenges have led to extensive research on modified theories of gravity \cite{Sotiriou:2008rp, Clifton:2011jh, Joyce:2014kja, Nojiri:2017ncd, Biswas:2005qr}, which may produce observable deviations from GR in strong-field or cosmological regimes while remaining consistent with current local and astrophysical observations. Regular black hole models have attracted considerable attention as potential resolutions to the pathologies present in classical black hole spacetimes.  In this context, several exact regular black hole solutions have been constructed within GR coupled to nonlinear electrodynamics (NED), particularly regarding configurations with electric or magnetic charges~\cite{Bronnikov:2000vy, Fan:2016hvf, Toshmatov:2018cks}. Research on regular black holes has also been extended to modified theories of gravity and their relationship with energy conditions~\cite{Capozziello:2013vna, Capozziello:2014bqa}. For example, regular solutions within four-dimensional $f(R)$ gravity coupled to non-linear electromagnetic fields have been obtained~\cite{Rodrigues:2015ayd}. Further regular black hole geometries have been derived through incorporating both electric and magnetic charges~\cite{Rodrigues:2018bdc}, introducing rotation~\cite{Bambi:2013ufa, Neves:2014aba, Toshmatov:2014nya, Azreg-Ainou:2014pra, Dymnikova:2015hka}, or investigating alternative gravitational theories~\cite{Berej:2006cc, Junior:2015fya,deSousaSilva:2018kkt, Rodrigues:2019xrc, Junior:2020zdt, Cano:2020ezi}.

Among the most compelling alternatives are black-bounce spacetimes and wormhole geometries, which resolve the classical curvature singularities present in GR. The black-bounce spacetimes, first introduced by Simpson and Visser \cite{Simpson:2018tsi}, provide a particularly elegant framework by offering a continuous interpolation between Schwarzschild black holes and traversable wormholes through a deformation parameter that regularizes the central singularity \cite{Simpson:2019mud,  Lobo:2017cay}. These spacetimes have attracted considerable attention because they include both regular black holes with event horizons and horizonless ultracompact objects, each with potentially distinct observational signatures \cite{Simpson:2019cer, Mazza:2021rgq, Franzin:2021vnj, Islam:2021ful}. The non-static extension of these black-bounce geometries was subsequently investigated using Eddington--Finkelstein coordinates~\cite{Simpson:2019cer}. In this dynamical context, the spacetime interpolates among Vaidya geometries, black-bounces, and traversable wormholes, enabling the modeling of processes such as accreting black-bounces, transitions from wormholes to black-bounces, and the reverse transition from black-bounces to wormholes.  The framework was later generalized by Lobo \textit{et al.} \cite{Lobo:2020ffi} by introducing a broader class of black-bounce spacetimes with more general mass functions, allowing a wider study of regular compact objects.
Franzin \textit{et al.}~\cite{Franzin:2021vnj} subsequently analyzed the spacetime structure of charged black-bounce Reissner--Nordström and black-bounce Kerr--Newman solutions. The Simpson--Visser black-bounce model~\cite{Simpson:2018tsi} was later identified as a specific case within this broader family~\cite{Lobo:2020ffi}.  The black-bounce spacetimes described by Lobo \textit{et al.} are distinguished by a nonvanishing area radius at all points, resulting in a throat that may be timelike, spacelike, or null. These cases correspond to traversable wormholes, bounces into a future universe, or one-way wormholes, respectively.

Strong gravitational lensing \cite{Nascimento:2020ime,Ovgun:2020yuv,Tsukamoto:2020bjm,Cheng:2021hoc,Islam:2021ful,Ditta:2023aw,Ditta:2023rhr,Feng:2024jeq,Ashraf:2025cnz} and shadows \cite{Bronnikov:2021liv,Guerrero:2021ues}  have been widely studied for both static and rotating solutions. Nascimento \textit{et al.} \cite{Nascimento:2020ime} generalized the lensing analysis to a broader class of black-bounce spacetimes. Cheng and Xie \cite{Zhang:2022nnj} considered the charged black-bounce--Reissner--Nordström spacetime, while Ghosh and Bhattacharyya \cite{Ghosh:2022mka} further extended the study to the charged rotating Kerr--Newman black-bounce spacetime. These studies have shown that such spacetimes can possess multiple photon spheres and antiphoton spheres, which can noticeably affect the lensing properties \cite{Cunha:2018acu, Perlick:2021aok}. In particular, horizonless configurations can contain both an unstable outer photon orbit and a stable inner photon orbit, producing image patterns that differ significantly from those of classical black holes \cite{Shaikh:2019itn, Tsukamoto:2020bjm}.

More importantly, Tsukamoto \cite{Tsukamoto:2020bjm} investigated strong gravitational lensing by the Simpson-Visser spacetime in a strong deflection limit for all the nonnegative parameters $a$ and $m$. In particular, when $a=3m$, the two photon spheres and the antiphoton sphere at the throat merge into a marginally unstable photon sphere. The deflection angle of light rays reflected by this photon sphere shows a non-logarithmic divergence in the strong deflection limit. Islam \textit{et al.} \cite{Islam:2021ful} extended this work to rotating Simpson--Visser black holes, showing how rotation changes the photon orbits and the corresponding lensing observables.  Recent work by Shaikh \cite{Shaikh:2019itn} introduced a gravitational lensing scenario in which light rays traverse the photon sphere, reflect near the effective potential barrier of the antiphoton sphere, and ultimately reach a distant observer. This analysis considered a general asymptotically flat, static, and spherically symmetric ultracompact object without an event horizon~\cite{Shaikh:2019itn}. Multiple studies have investigated the observational signatures of spacetimes with multiple photon spheres. Specifically, Shaikh \textit{et al.} analyzed gravitational lensing in reflection-symmetric wormhole geometries containing a photon sphere at the throat and an additional photon sphere away from the throat~\cite{Shaikh:2018oul,Shaikh:2019jfr,Godani:2021atr}. Additionally, Gan \textit{et al.} studied the shadow properties of hairy black holes possessing two photon spheres~\cite{Gan:2021pwu,Gan:2021xdl}. Reflection-asymmetric thin-shell wormholes with two photon spheres of unequal radii have also been proposed; these configurations can produce shadow images accompanied by light rings~\cite{Wang:2020emr,Wielgus:2020uqz,Tsukamoto:2021fpp,Guerrero:2021pxt,Peng:2021osd}.

Numerous properties of photon spheres have been extensively studied, including bounds on their radii~\cite{Hod:2017xkz, Peng:2018nkj, Hod:2020pim, Hod:2023jmx}, the possible number of photon spheres~\cite{Hod:2017zpi, Cunha:2020azh, Cunha:2024ajc, Padhye:2024yrj}, and their relationships to photon absorption cross sections~\cite{Sanchez:1977si, Decanini:2010fz, Wei:2011zw}, quasinormal modes~\cite{Press:1971wr, Stefanov:2010xz, Raffaelli:2014ola, Igata:2025plb}, and the stability of thin-shell wormholes~\cite{Barcelo:2000ta, Koga:2020gqd, Tsukamoto:2023kvk}. The study of shadow formation and strong-field gravitational lensing in spacetimes with multiple photon spheres is a recent development, leaving many aspects of these phenomena unexplored across the full parameter space of relevant models. Particularly, the horizonless objects which have received significant attention for various reasons (see \cite{Cardoso:2016rao,Cunha:2018gql,Carballo-Rubio:2018jzw,Konoplya:2018yrp}). Considerable research has also focused on determining whether, and to what extent, these horizonless compact objects can be distinguished from black holes. Furthermore, future space-based observational missions are expected to detect photons that are strongly bent near these photon spheres~\cite{Lupsasca:2024xhq,Johnson:2024ttr}. Therefore, investigating the behavior of light trajectories near photon spheres is crucial not only for black hole geometries but also for alternative compact configurations, including horizon-less compact objects such as wormholes and naked singularities~\cite{Shaikh:2019itn,Shaikh:2018lcc,Virbhadra:2002ju,Virbhadra:1998dy,Zhang:2024sgs,Solodukhin:2025opw,Olmo:2023lil,Shaikh:2019jfr,Bronnikov:2018nub,Shaikh:2018oul,Shaikh:2018kfv,Tsukamoto:2016qro,Tsukamoto:2016zdu,Tsukamoto:2012xs,Nandi:2006ds}. 
A comprehensive analysis of the deflection angle in the strong deflection regime, particularly over a broad range of spacetime parameters for rotating spacetimes, is therefore necessary. Such investigations can identify distinctive observational signatures and differentiate the shadow and lensing features of horizonless objects with multiple photon spheres from those that possess a single photon sphere.

In this work, we extend previous studies of gravitational lensing in black-bounce spacetimes to two rotating black-bounce models proposed by Lobo \textit{et al.} \cite{Lobo:2020ffi}. These models generalise the Simpson--Visser spacetime by using different mass functions, leading to distinct near-throat geometries and photon-sphere structures. We derive the strong-deflection angles for both black-hole and horizonless solutions and calculate the corresponding lensing observables for photon trajectories associated with the outer and inner photon spheres. The predicted shadow angular diameters of the models are compared with the EHT measurement of Sgr A* to constrain the model parameters. We also discuss whether these spacetimes can be distinguished from the Kerr solution with future observations by the next-generation EHT (ng-EHT) \cite{Ayzenberg:2023hfw, Lupsasca:2024xhq}.

This paper is organized as follows. In Sec.~\ref {Sec2} and \ref{sec3}, we review the general black-bounce spacetimes and construct their rotating counterparts using the modified Newman--Janis algorithm. In Sec.~\ref {sec4}, we analyse the geodesic structure and unstable photon orbits in these rotating black-bounce geometries. We lay the foundation for Strong gravitational lensing in the given black-bounce models in Sec.~\ref {sec5} and investigate the deflection angle in the strong-deflection regime in Sec.~\ref {sec6}. We examine the corresponding gravitational lensing observables  both inside and outside the photon sphere in Sec.~\ref{sec7}. Throughout the analysis, we assume that both the source and the observer are located on the opposite sides of the lens and are situated far from the photon spheres. Finally, in Sec.~\ref{sec8}, we conclude with a summary of the results.   Unless otherwise stated, we adopt the units such that speed of light and Newton's gravitational constant set to unity, i.e., $c=G=1$.

\section{General black-bounce spacetimes}\label{Sec2}
Regular black holes and regular black-bounce spacetimes have emerged as promising alternatives to classical GR singular black-hole solutions, providing spacetime models in which curvature invariants remain finite throughout.  \cite{Bardeen:1968,Hayward:2005gi,AyonBeato:2000zs,AyonBeato:1998ub}. In particular, black-bounce spacetimes constitute a broad class of regular geometries that smoothly interpolate between regular black holes and traversable wormholes, depending on the choice of parameters \cite{Simpson:2018tsi,Lobo:2020ffi,Franzin:2021vnj}. Owing to their rich causal structure and absence of curvature singularities, these solutions have been extensively investigated in the contexts of black-hole shadows \cite{KumarWalia:2022aop,Kumar:2020yem}, quasinormal modes \cite{Yang:2022xxh}, accretion processes \cite{Bambhaniya:2021ugr}, gravitational lensing \cite{Kumar:2022fqo},  and other applications \cite{Simpson:2018tsi,Lobo:2020ffi,Franzin:2021vnj,Tzikas:2018cvs,Bronnikov:2024izh, Amir:2015pja,Alencar:2024yvh}.
The general black-bounce framework introduced by Lobo \textit{et al.} \cite{Lobo:2020ffi} encompasses several regular black-hole and wormhole configurations within a unified geometric description. Before constructing the rotating counterparts and analyzing their optical properties, it is useful to review the underlying static geometry and the conditions required for spacetime regularity. These conditions can be conveniently expressed in terms of curvature invariants, which remain finite throughout the spacetime and thereby guarantee the absence of curvature singularities.
\subsection{Metric and curvature}\label{SS:metric}
A general static and spherically symmetric spacetime may be written as
\begin{eqnarray}\label{metric1}
ds^2 & =-\left(1-\frac{2M(r)}{\Sigma(r)}\right)dt^2+\frac{dr^2}{\left(1-\frac{2M(r)}{\Sigma(r)}\right)} \nonumber \\ 
& +\Sigma^2(r)\left(d\theta^2+\sin^2\theta d\phi^2\right),
\end{eqnarray}
where $M(r)$ and $\Sigma(r)$ are arbitrary functions characterizing the mass profile and areal radius of the geometry, respectively. The metric determinant is
\begin{equation}
g=-\Sigma^4(r)\sin^2\theta,
\end{equation}
while the area of a two-sphere of constant radial coordinate $r$ is
\begin{equation}
\bar{A}(r)=4\pi\Sigma^2(r).
\end{equation}
The line element (\ref{metric1}) provides a convenient starting point for constructing a broad class of regular compact-object geometries, including regular black holes, black bounces, and traversable wormholes \cite{Visser:1995cc,Simpson:2018tsi,Lobo:2020ffi}.
To investigate the regularity properties of the spacetime, we consider the Kretschmann scalar
$K=R_{\alpha\beta\mu\nu}R^{\alpha\beta\mu\nu}$ and the Weyl invariant
$C=C_{\mu\nu\alpha\beta}C^{\mu\nu\alpha\beta}$. For the metric (\ref{metric1}), these invariants take the form
\begin{widetext}
\begin{eqnarray}
K =
\frac{4}{\Sigma^6}
\Big[
2 \Sigma'^2 \mathcal{A}^2
+2 \big(\Sigma'\mathcal{A}
+\Sigma \mathcal{B}\Sigma''\big)^2
+\big(\Sigma \mathcal{C}
+2M\Sigma'^2\big)^2
+\big(\mathcal{B}\Sigma'^2+\Sigma\big)^2
\Big],
\label{Kret}
\
C =
\frac{4}{3\Sigma^6}
\Big[
\Sigma \mathcal{C}
+3\mathcal{B}\Sigma'^2
-\Sigma\Sigma'^2
+\Sigma
\Big]^2,
\label{Csq}
\end{eqnarray}
\end{widetext}
where a prime denotes derivative with respect to $r$. The functions $\mathcal{A}(r)$, $\mathcal{B}(r)$, and $\mathcal{C}(r)$ are defined below. The spacetime is regular provided $\Sigma(r)\neq0$ everywhere and both curvature invariants (\ref{Kret}) and (\ref{Csq}) remain finite throughout, ensuring the absence of curvature singularities and constituting the black-bounce and regular black-hole geometries \cite{Bardeen:1968,Hayward:2005gi,Simpson:2018tsi,Lobo:2020ffi,Franzin:2021vnj}.
\begin{align}
\mathcal{A}(r) &\equiv \Sigma M' - M \Sigma' ,\\
\mathcal{B}(r) &\equiv 2M - \Sigma ,\\
\mathcal{C}(r) &\equiv \Sigma M'' - 2M'\Sigma' - M\Sigma'' .
\end{align}
Regularity of the static spacetime requires $\Sigma(r)\neq0$ and the finiteness of the scalars \eqref{Kret}  for all $r$.   

\subsection{Rotating Metric}
The rotating metric in Eq.~\eqref{metric2} can be obtained via the modified Newman–Janis algorithm (MNJA), which generates axially symmetric spacetimes from a given \emph{static seed} black hole metric.  Originally, the NJA algorithm was introduced to recover the Kerr solution from Schwarzschild~\cite{Newman:1965tw}, the standard NJA algorithm  relies on a complexification step whose applicability becomes ambiguous for non-vacuum or regular spacetimes.
The MNJA proposed by Azreg-Aïnou~\citep{Azreg-Ainou:2014aqa,Azreg-Ainou:2014pra} circumvents this issue by avoiding complex coordinate transformations and working instead with a real tetrad reconstruction. This makes the procedure particularly suitable for constructing rotating extensions of regular black holes, black-bounce spscetimes~\cite{Jusufi:2019caq,Azreg-Ainou:2014aqa,Azreg-Ainou:2014pra,Mazza:2021rgq} and in modified theories of gravities \cite{Kumar:2020owy,Kumar:2020hgm,Ghosh:2014pba,Ghosh:2021clx,Ali:2024ssf,Kumar:2022vfg} . Applying it to the seed metric~\eqref{metric1} leads to the Kerr-like form~\eqref{metric1}, where the functions $M(r)$ and $\Sigma(r)$ encode the deviation from Kerr in the non-rotating limit.
The rotating counterpart of the spherically symmetric spacetime metric~\eqref{metric1} in Boyer–Lindquist coordinates can be written in a Kerr-like form as
 \begin{eqnarray}\label{metric2}
ds^2 &&=
- \left(1 - \frac{2 M(r)\Sigma(r)}{\rho^2} \right) dt^2
+ \frac{\rho^2}{\Delta} dr^2
+ \rho^2 \, d\theta^2 \nonumber \\ && 
- \frac{4 a M(r)\Sigma(r)}{\rho^2} \sin^2\theta \, dt \, d\phi
+ \frac{\mathcal{H} \sin^2\theta}{\rho^2} \, d\phi^2\nonumber\\
\Delta &&= r^2+a^2-2M(r)\Sigma(r)\nonumber\\
\mathcal{H} &&=  \left(r^2+ a^2 +
\frac{2M(r)\Sigma(r)  a^2 }{\rho^2} \sin^2 \theta
\right)\nonumber\\
\rho^2 &&= \Sigma(r)^2 + a^2 \cos^2 \theta
\end{eqnarray}
 where $a$ is the rotation parameter. This form generalises the Kerr metric~\cite{Kerr:1963ud}. 
For $M(r)=M$ and $\Sigma(r)=r$, the above metric reduces to the Kerr solution ~\cite{Kerr:1963ud} in Boyer–Lindquist coordinates. More generally, suitable choices of $M(r)$ and $\Sigma(r)$ generate rotating regular black holes and black-bounce geometries~\cite{Hayward:2006ej,Simpson:2018tsi,Simpson:2019mud,Modesto:2017sdr,Bronnikov:2021uta}. A representative example is the Simpson–Visser black-bounce spacetime~\cite{Simpson:2018tsi,Simpson:2019mud}, obtained by choosing
 \begin{eqnarray}\label{eqn1}
 \Sigma(r)=\sqrt{r^2+\ell^2}, \qquad M(r)=M,
 \end{eqnarray}
 where $\ell$ is the bounce parameter controlling the regularisation of the central region. The spacetime is everywhere regular and interpolates between a black hole, a one-way wormhole, and a two-way traversable wormhole depending on the parameter range.
The horizons of the metric (\ref{metric2}) are determined by zeros of $\Delta=0$. For $|a|<M$,  $\Delta=0$ admits two positive real roots corresponding to two horizons when $\ell < M-\sqrt{M^2-a^2}$, while a single degenerate horizon occurs at $\ell = M-\sqrt{M^2-a^2}$. We find that no horizons exists for $\ell > M-\sqrt{M^2-a^2}$, and when $\ell > M+\sqrt{M^2-a^2}$ the spacetime  metric (\ref{metric2}) describes a two-way traversable wormhole. The limiting values $\ell = M \pm \sqrt{M^2-a^2}$ correspond to null-throat configurations, with the former describing an extremal black hole and the latter a one-way traversable wormhole.  The extremal case $|a|=M$, degenerate horizons exist only for $\ell<M$, while the spacetime becomes horizonless for $\ell>M$. For $|a|>M$, no horizons exist for any value of $\ell$, yielding a regular over-spinning geometry.

\begingroup
\begin{table*}
	\caption{Summary of the three black-bounce spacetimes analyzed in this work, listing the corresponding mass functions $M(r)$ and metric functions $f(r)$. The parameter $\ell$ characterizes the bounce scale and determines the transition between black hole and wormhole geometries.}\label{table1}
	\begin{ruledtabular}
    \renewcommand{\arraystretch}{2}
		\begin{tabular}{l c c c}
Spacetime & Value of $n$ and  $k$ & $M(r)$ & $f(r)$ \\
\hline
Model-I     & $n=2$~~$k=0$    & $\displaystyle \frac{M~\sqrt{r^2+\ell^2}}{\sqrt[4]{r^4+\ell^4}} $ & $\displaystyle 1-\frac{2M}{\sqrt[4]{r^4+\ell^4}}$ \\[10pt]

\hline
Model-II     & $n=1$~~$k=2$    & $\displaystyle  \frac{M r^2}{r^2 + \ell^2 }$ & $\displaystyle 1-\frac{2Mr^2}{(r^2+\ell^2)^{3/2}}$ \\[10pt]

\hline
Model-III     & $n=3/2$~~$k=2$    & $\displaystyle  \frac{M r^2 \sqrt{r^2 + \ell^2}}{r^3 + \ell^3 }$  & $\displaystyle 1-\frac{2Mr^2}{(r^3+\ell^3)}$ \\[10pt]
		\end{tabular}
	\end{ruledtabular}
\end{table*}
\endgroup

\section{New black-bounce spacetimes}\label{sec3}

The black-bounce geometries include a class of regular spacetimes in which a minimal-area surface replaces the central singularity, yielding either a regular black hole with a hidden throat or a traversable wormhole connecting two asymptotically flat regions. The black bounce spacetime provides a unified description interpolating between regular black holes and wormhole geometries~\cite{Simpson:2018tsi, Simpson:2019mud,Hayward:2006ej,Bronnikov:2021uta,Visser:1995cc}.
A more general class of black-bounce solutions extending the Simpson--Visser spacetime has been proposed in Ref.~\cite{Lobo:2020ffi}, where the metric functions are parametrised as
\begin{eqnarray}\label{newBBS}
M(r)=\frac{M\,\Sigma(r)\, r^{k}}{\left(r^{2n}+\ell^{2n}\right)^{(k+1)/(2n)}}
\end{eqnarray}
Here $n$ and $k$ are positive integers controlling the radial fall-off and deformation of the effective mass function, while $\ell$ denotes the bounce scale governing the regularisation of the central region. The Simpson--Visser spacetime is recovered for $n=1$ and $k=0$~\cite{Simpson:2018tsi,Simpson:2019mud}. In the static limit $a\to 0$, the spacetime reduces to a spherically symmetric configuration that approaches Schwarzschild asymptotically for $\ell\to 0$.

It is worth noting that, due to the specific choice $\Sigma(r)=\sqrt{r^{2}+\ell^{2}}$, this class does not reproduce standard regular black-hole solutions such as the Bardeen~\cite{Bardeen:1968,Bambi:2013ufa}, Hayward~\cite{Hayward:2006ej,Modesto:2004xx}, or Frolov-type geometries~\cite{Frolov:2016pav}. These arise from different effective matter models or nonlinear electrodynamics sources. Nevertheless, the present construction generates a broad family of black-bounce spacetimes with distinct near-throat behaviors, some representative cases being summarized in Table~\ref{table1}. In particular, \textit{Model-I} reproduces the qualitative causal structure of the Simpson--Visser spacetime~\cite{Lobo:2020ffi}, and for $k=0$ (arbitrary $n$) the global structure remains unchanged.

The lensing properties of the Simpson--Visser spacetime have been extensively studied in Refs.~\cite{Tsukamoto:2020bjm,Islam:2021ful,Ovgun:2020yuv}. More generally, strong-deflection gravitational lensing in regular and black-bounce geometries has been analyzed in Refs.~\cite{Perlick:2004tq,Tsukamoto:2020bjm, Nascimento:2020ime,Ghosh:2022mka}. The deformations introduced in \textit{Model-II} and \textit{Model-III} modify the effective mass profile and, consequently, the photon-sphere structure and strong-lensing observables. The main goal of this work is to systematically investigate the strong gravitational lensing signatures of the rotating extensions of these geometries and to identify possible deviations from the Kerr predictions.

\begin{figure*}
     \includegraphics[scale=0.65]{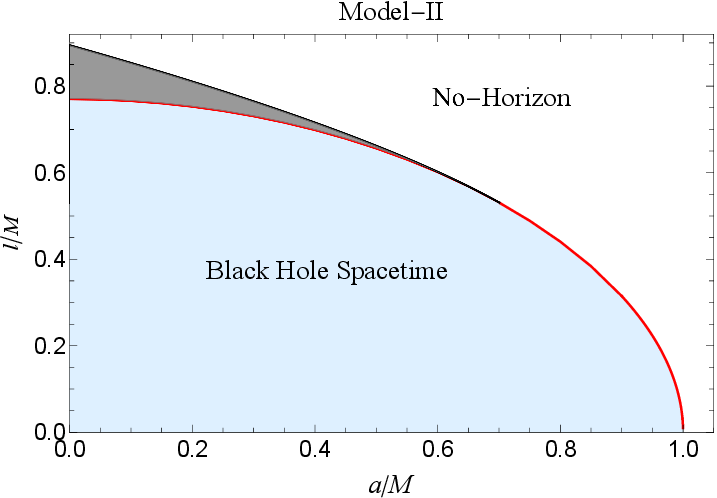}
     \hspace{0.8cm}
    \includegraphics[scale=0.65]{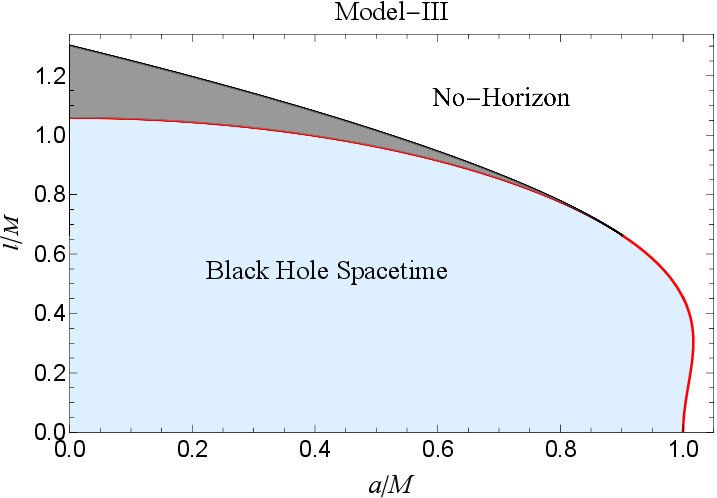}
    \caption{Parameter space in the $(a, l)$ plane illustrating the distinction between black hole and no-horizon spacetimes for \textit{Model-II} and \textit{Model-III}. The light blue regions correspond to parameter values that admit a black hole solution, while regions outside the red boundary curve represent no-horizon spacetimes. Among these, only the gray shaded subset yields a closed photon ring structure.
}\label{parspace}
\end{figure*}

\section{Geodesics and unstable photon orbits in black-bounce spacetimes}\label{sec4}
Photon geodesics in the vicinity of compact objects play a central role in probing strong-field gravity and its observational signatures, including black hole shadows, photon rings, and relativistic lensing images~\cite{Bardeen:1972fi,Falcke:1999pj,Perlick:2004tq,Bozza:2002zj,Virbhadra:1999nm,Grenzebach:2015oea}. In particular, unstable circular null orbits (photon spheres) encode key geometric properties of the spacetime and determine the structure of the shadow boundary.
To study null geodesics in the rotating black-bounce spacetime~(\ref{metric2}), we employ the Hamilton--Jacobi formalism~\cite{Carter:1968rr,Chandrasekhar:1985kt}. The Hamilton--Jacobi equation is given by
\begin{eqnarray}\label{HmaJam}
\frac{\partial S}{\partial \tau}
= -\frac{1}{2} g^{\mu\nu}
\frac{\partial S}{\partial x^\mu}
\frac{\partial S}{\partial x^\nu},
\end{eqnarray}
where $\tau$ is an affine parameter along null geodesics and $S$ is the Jacobi action.

The spacetime metric (\ref{metric2}) is  stationarity and axial symmetry,  imply the existence of Killing vectors $\chi^\mu_{(t)}$ and $\chi^\mu_{(\phi)}$, leading to conserved quantities: the energy ${\cal E}=-\chi^{\mu}_{(t)}p_{\mu}$ and the azimuthal angular momentum ${J}=\chi^{\mu}_{(\phi)}p_{\mu}$. Following Carter’s separability ansatz~\cite{Carter:1968rr}, we can write the separable action as
\begin{eqnarray}
S = -{\cal E} t + {J}\phi + S_r(r) + S_\theta(\theta),
\label{action}
\end{eqnarray}
where $S_r(r)$ and $S_\theta(\theta)$, respectively, depend only on $r$ and $\theta$ and separability introduces the Carter constant ${\cal K}$.

Using Eqs.~(\ref{HmaJam}) and (\ref{action}), the geodesic equations take the standard first-order form~\cite{Carter:1968rr,Chandrasekhar:1985kt,Teo:2003ltt}
\begin{align}
\rho^2 \frac{dt}{d\tau} &=
\frac{\Sigma^2+a^2}{\Delta(r)}
\left[{\cal E}(\Sigma^2+a^2)-a{J}\right]
- a\left(a{\cal E}\sin^2\theta - {J}\right),
\label{tuch}\\[2mm]
\rho^2 \frac{dr}{d\tau} &= \pm \sqrt{\mathcal{R}_r(r)},
\label{r}\\[2mm]
\rho^2 \frac{d\theta}{d\tau} &= \pm \sqrt{\Theta_\theta(\theta)},
\label{th}\\[2mm]
\rho^2 \frac{d\phi}{d\tau} &=
\frac{a}{\Delta(r)}
\left[{\cal E}(\Sigma^2+a^2)-a{J}\right]
- \left(a{\cal E}-\frac{J}{\sin^2\theta}\right),
\label{phiuch}
\end{align}
with radial and polar potentials
\begin{eqnarray}
\mathcal{R}_r(r)
&=& \left[(\Sigma^2+a^2){\cal E}-a{J}\right]^2
- \Delta(r)\left[{\cal K}+(a{\cal E}-{J})^2\right]\quad\quad
\label{radialP}
\\
\Theta_\theta(\theta)
&=& {\cal K}
- \left(\frac{J^2}{\sin^2\theta}-a^2 {\cal E}^2\right)\cos^2\theta.
\label{theta0}
\end{eqnarray}
The conditions $\mathcal{R}_r(r)\geq 0$ and $\Theta_\theta(\theta)\geq 0$ define the allowed regions of motion. Unstable circular photon orbits are determined by the conditions $\mathcal{R}_r=0$ and $d\mathcal{R}_r/dr=0$, which define the photon sphere structure governing shadow formation and strong-field lensing observables~\cite{Synge:1966okc,Bardeen:1973tla,Perlick:2004tq,Grenzebach:2014fha,Johannsen:2010ru}.

Here, $\mathcal{K}$ is the Carter constant of motion and controls how photons move in the polar direction. When $\mathcal{K}=0$, the motion in the $\theta$ direction vanishes, and photons remain confined to the equatorial plane $\theta=\pi/2$. In this case, circular photon orbits can exist~\cite{Carter:1968rr,Chandrasekhar:1985kt}.
In black hole spacetimes, photons with sufficient energy relative to the effective radial potential can overcome the potential barrier and fall through the event horizon. These photons do not return and contribute to the dark region seen in the black hole shadow~\cite{Synge:1966okc,Bardeen:1973tla}. In contrast, in horizonless regular spacetimes, there is no event horizon to absorb the photons. Instead, the effective potential near the center acts as a barrier, pushing photons back and allowing them to escape to a distant observer. This change strongly affects the observed image and lensing pattern~\cite{Perlick:2004tq,Bozza:2002zj}.

Besides unstable circular photon orbits, horizonless spacetimes can also allow stable circular photon orbits. This happens because the radial potential $\mathcal{R}_r(r)$ may have both a maximum and a minimum. Photons with energy exactly equal to the peak of the potential stay on circular orbits and mark the transition between falling in and escaping. These orbits are determined by~\cite{Chandrasekhar:1985kt,Teo:2003ltt}
\begin{equation}
\left.\mathcal{R}_r\right|_{r=r_m}=0, 
\qquad
\left.\frac{d\mathcal{R}_r}{dr}\right|_{r=r_m}=0,
\end{equation}
where $r_m$ is the radius of the circular photon orbit.

The stability of circular photon orbits is governed by second derivative of the radial effective potential. These orbits correspond to stationary points of $\mathcal{R}_r(r)$, and their stability is determined by the second derivative of the potential in the radial direction~\cite{Carter:1968rr,Chandrasekhar:1985kt}. If
\begin{equation}
\left.\frac{d^2 \mathcal{R}_r}{dr^2}\right|_{r=r_m} < 0,
\end{equation}
the circular orbit lies at a local maximum of the effective potential, implying that small radial perturbations grow exponentially. Such orbits are therefore unstable and form the photon sphere that governs the boundary of the black hole shadow. In contrast,  if
\begin{equation}
\left.\frac{d^2 \mathcal{R}_r}{dr^2}\right|_{r=r_m} > 0,
\end{equation}
the orbit corresponds to a local minimum and is stable. The case of equality represents the transition point between stability and instability~\cite{Carter:1968rr,Chandrasekhar:1985kt,Teo:2003ltt,Grenzebach:2014fha}.
Hereafter, we define a photon sphere as a sphere 
filled with unstable circular light orbits, while an antiphoton sphere 
corresponds to stable circular light orbits.

\begin{figure*}
    \begin{tabular}{c c c}
     \includegraphics[scale=0.55]{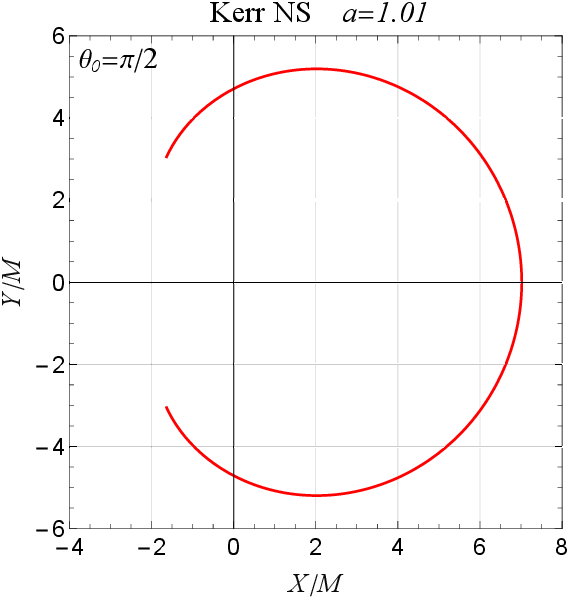}&
     \includegraphics[scale=0.55]{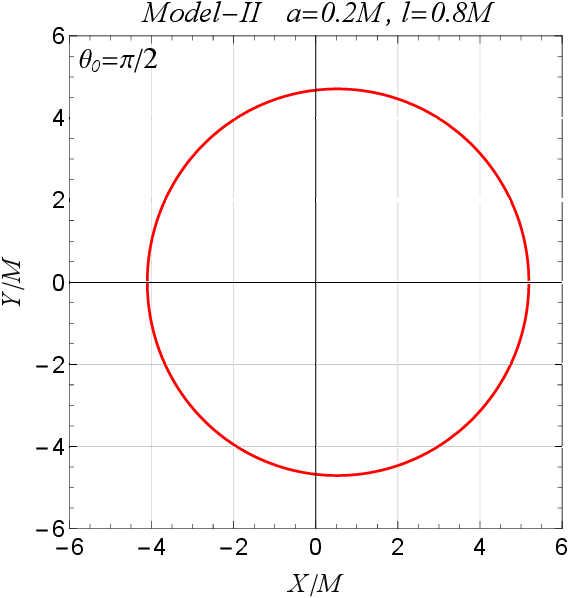}&
     \includegraphics[scale=0.55]{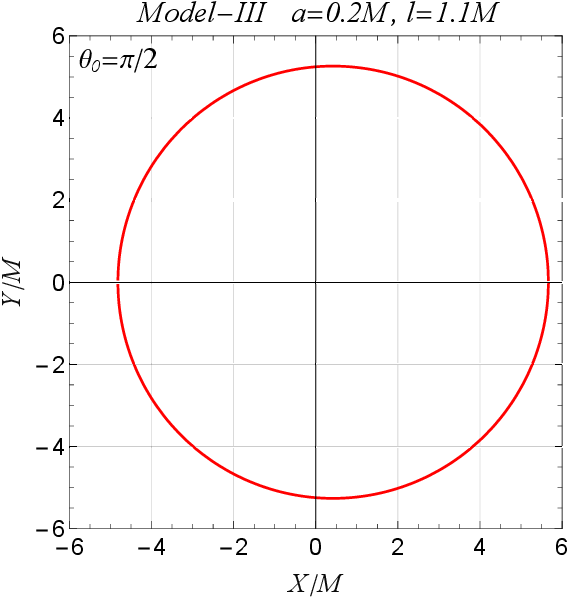}
    \end{tabular}
    \caption{Comparison of photon rings for rotating black bounce models and the Kerr naked singularity (NS) at an inclination angle $\theta_0 = \pi/2$. The Kerr naked singularity  produces an open arc–like photon ring, in contrast to the no-horizon spacetimes of the black bounce models, which exhibit closed photon ring structures.}\label{photonring}
\end{figure*}

\begin{figure*}
     \includegraphics[scale=0.6]{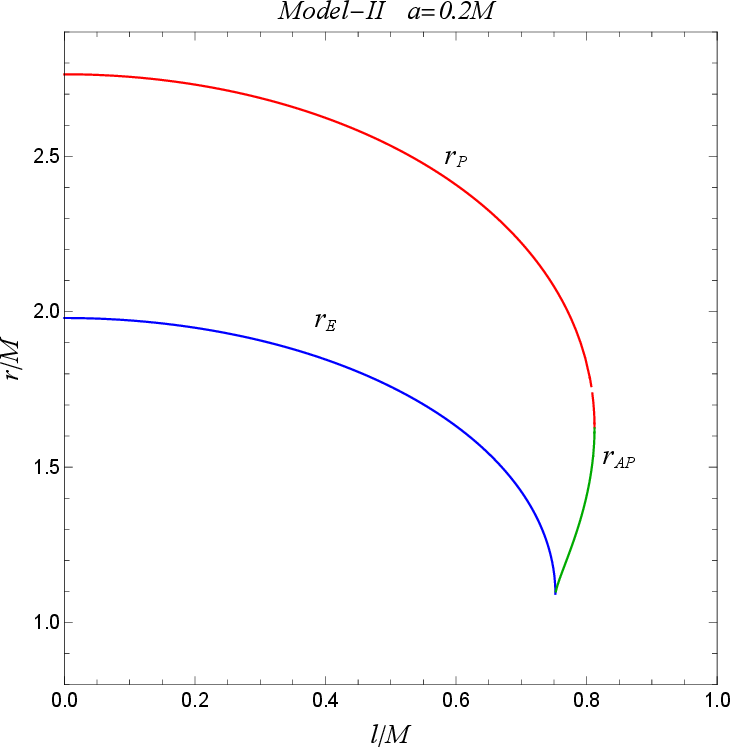}
     \hspace{1.5cm}
    \includegraphics[scale=0.6]{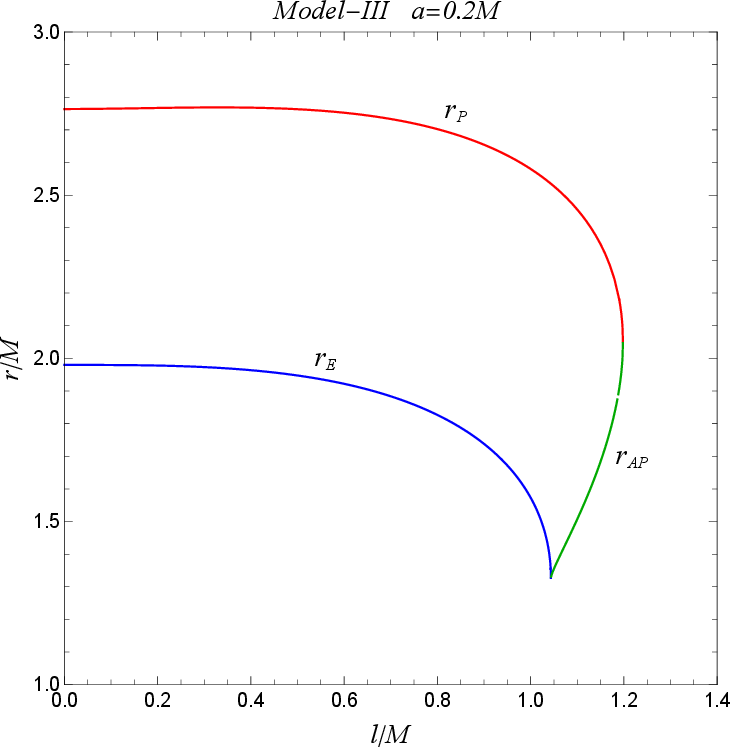}
    \caption{Reduced radial coordinates of the circular prograde photon orbit radii $(r^{-}_{P}/M)$, antiphoton sphere $(r_{ap}/M)$, and event horizon $(r_{E}/M)$ at $(a = 0.2M)$ shown respectively by the red, green, and blue curves for \textit{Model-II} and \textit{Model-III}. For \textit{Model-II}, the the photon orbit radii spans  $(1.628M \leq r^{-}_{P} \leq 2.763M)$ for $(0 \leq \ell \leq 0.812M)$, the antiphoton sphere lies within $(1.1003M \leq r_{ap} \leq 1.62443M)$ for $(0.7523M \leq \ell \leq 0.812M)$, while the event horizon covers $( 1.09262M \leq r_{E} \leq 1.9798M)$ for $(0 \leq \ell \leq 0.7523M)$. Similarly, in \textit{Model-III}, the the photon orbit radii is between  $(2.05M \leq r^{-}_{P} \leq 2.76387)$ for $(0 \leq \ell \leq 1.198M)$, the antiphoton sphere lies within $(1.33M \leq r_{ap} \leq 2.046M)$ for $(1.043M \leq \ell \leq 1.198M)$, while the event horizon covers $(1.327M \leq r_{E} \leq 1.979M)$ for $(0 \leq \ell \leq 1.0435M)$.
}\label{photonOrbit}
\end{figure*}

\subsection{Model-II}
For \textit{Model-II}, the mass function listed in Table~\ref{table1} yields the metric function
\begin{equation}\label{Delta1}
\Delta(r)=r^{2}+\ell^{2}+a^{2}
-\frac{2Mr^{2}}{\sqrt{r^{2}+\ell^{2}}}.
\end{equation}
The horizon of the spacetime are the roots of $\Delta(r)=0$. The equation admits two positive and two negative real roots for $0<\ell<\ell_{-}$, a pair of degenerate roots at $\ell=\ell_{-}$, and no real roots for $\ell>\ell_{-}$, where
\begin{equation}\label{Eq10}
\ell_{-}=
\frac{1}{3\sqrt{3}}
\sqrt{\frac{(4M^{2}-3a^{2})^{3/2}}{M}
+8M^{2}-9a^{2}}.
\end{equation}
The positive roots correspond to Killing horizons, while the negative roots arise from the extension of the radial coordinate through the bounce. For $\ell>\ell_{-}$, the spacetime becomes horizonless and possesses a regular traversable throat at $r=0$, characteristic of black-bounce geometries that smoothly interpolate between regular black holes and wormholes~\cite{Simpson:2018tsi,Simpson:2019mud,Lobo:2020ffi,Bronnikov:2021uta}.
For the parameter space, we set $M=1$. For a fixed spin parameter $a$, there exists a critical value $\ell=\ell_E$ at which $\Delta(r)=0$ admits a double root at $r=r_H$, which corresponds to an extremal configuration where the inner and outer horizons merge, analogous to the extremal Kerr black holes~\cite{Kerr:1963ud,Chandrasekhar:1985kt}. For $\ell>\ell_E$, no real roots of $\Delta(r)$ exist and the spacetime becomes completely horizonless.
The resulting parameter space is depicted in Fig.~\ref{parspace} with the light-blue region corresponds to black-hole solutions possessing one or two horizons, while the red boundary curve marks the extremal where two horizons coincides. Beyond this boundary, the spacetime is horizonless or no-horizon spacetimes. Unlike the Kerr naked singularity, however, the horizonless sector of the black-bounce spacetime exhibits a distinct gray region bounded by $\ell_E<\ell<\ell_P$, within which closed photon-ring structures still possible. This behavior reflects the regular nature of the geometry and the existence of nontrivial photon trapping regions even for no-horizon spacetime~\cite{Cunha:2017qtt,Cardoso:2014sna,Kumar:2020yem, Grenzebach:2014fha, Kumar:2020ltt}.

A representative example of such a closed photon ring for \textit{Model-II}, corresponding to $a=0.2M$ and $\ell=1.08M$, is shown in Fig.~\ref{photonring} and compared with the corresponding Kerr naked-singularity case. Only when $\ell>\ell_P$ (the white region in Fig.~\ref{parspace}) do the photon rings transition into open arc-like structures resembling those found in over-spinning Kerr geometries~\cite{Bambi:2010hf,Stuchlik:2019uvf}.

To illustrate the behavior of circular photon orbits, we consider $a=0.2M$. We find that a single unstable circular photon orbit exists throughout the black-hole regime, $0\le \ell \le \ell_E =0.7523M$. In contrast, within the horizonless interval $\ell_E \le \ell \le \ell_P =0.81M$, two circular photon orbits appear: an inner stable orbit and an outer unstable orbit, as shown in Fig.~\ref{photonOrbit}. Similar multiple-photon-orbit structures have previously been reported in horizonless compact objects and wormhole spacetimes~\cite{Cardoso:2014sna,Cunha:2018acu}. The event-horizon radius $r_E$, defined as the largest positive root of Eq.~(\ref{Delta1}), is also plotted in Fig.~\ref{photonOrbit} for comparison.

An interesting feature emerges for extremal configurations. For small values of $\ell$, the horizon radius coincides with the radius of the prograde circular photon orbit, as in the extremal Kerr geometry~\cite{Bardeen:1972fi,Teo:2003ltt}. As $\ell$ increases, however, the two radii separate, giving rise to a finite gap
\begin{equation}
\delta=\frac{r_m^{-}-r_E}{M}>0.
\end{equation}
This separation reflects the influence of the bounce parameter on the near-horizon geometry and provides a useful measure of the departure from the Kerr spacetime. Table~\ref{table1A} lists the extremal horizon radius together with the radii of the prograde and retrograde circular photon orbits, $r^{-}$ and $r^{+}$, for several representative extremal configurations characterized by $(\ell,a_E(\ell))$.

\subsection{Model-III}
In this model, $\Delta(r)$ takes the form: 
\begin{equation}\label{Delta2}
    \Delta(r) = r^2 +l^2 + a^2 - \frac{2 M (r^2 + l^2) r^2}{r^3 + l^3}
\end{equation}
\begingroup
\begin{table*}
	\caption{Summary of the parameter $(\ell)$ and the corresponding extremal values of spin $(a_E)$ for \textit{Model-II}. For each combination, the degenerate horizon radius $(r_{E})$ as well as the radii of the prograde $r_{m}^{-}$ and retrograde $r_{m}^{+}$ photon orbits are computed and listed.
}\label{table1A}
	\begin{ruledtabular}
		\begin{tabular}{l c c c c c c}
$\ell/M$&$a_{E}/M $ & $r_{E}/M $ & $r_{m}^{+}/M $ & $r_{m}^{-}/M $ &$\delta = (r_{m}^{-}-r_{E})/M $ \\
\hline
 0.   & 1.           & 1.0       & 4.0        & 1.00     & 0.0 \\
 0.1  & 0.989999     & 1.00498   & 3.97047   & 1.00504   & 0.0000610476 \\
 0.2  & 0.95994      & 1.01785   & 3.92889   & 1.01792   & 0.0000696341 \\
 0.3  & 0.909363     & 1.03504   & 3.85784   & 1.03513   & 0.0000898821 \\
 0.4  & 0.83662      & 1.05285   & 3.75350    & 1.053     & 0.000151465 \\
 0.5  & 0.737476     & 1.06856   & 3.60767   & 1.10338   & 0.0348164 \\
 0.6  & 0.601481     & 1.08042   & 3.40168   & 1.34716   & 0.266742 \\
 0.7  & 0.395251     & 1.08729   & 3.07769   & 1.72324   & 0.635952 \\
 0.72 & 0.335399     & 1.08799  & 2.98123   & 1.83147   & 0.743472 \\
 0.74 & 0.260625     & 1.08846  & 2.85926   & 1.96557   & 0.877106 \\
 0.76 & 0.150122    & 1.08869   & 2.67609   & 2.1612    & 1.07252 \\
0.7698& 0.0         & 1.08866   & 2.42175    & 2.42175  & 1.33309\\

		\end{tabular}
	\end{ruledtabular}
\end{table*}
\endgroup
\begingroup
\begin{table*}
	\caption{Summary of the parameter $(\ell)$ and the corresponding extremal values of spin $(a_E)$ for \textit{Model-III}. For each combination, the event horizon radius $(r_{E})$ as well as the radii of the prograde $r_{m}^{-}$ and retrograde $r_{m}^{+}$ photon orbits are computed and listed.
}\label{table1B}
	\begin{ruledtabular}
		\begin{tabular}{l c c c c c c}
$\ell/M$&$a_{E}/M $ & $r_{E}/M $ & $r_{m}^{+}/M $ & $r_{m}^{-}/M $ &$\delta = (r_{m}^{-}-r_{E})/M $ \\
\hline\hline
 0.0     &   1.0          & 1.0        & 4.0      & 1.0        & 0.0 \\
 0.1    &   1.00402     & 0.992065   & 3.98817  & 0.99393   & 0.0000582896 \\
 0.2    &   1.01195     & 0.976429   & 3.99604  & 0.978324   & 0.0000598402 \\
 0.3    &   1.01666     & 0.970234   & 4.00088  & 0.972192   & 0.0000686473 \\
 0.4    &   1.01026     & 0.987984   & 3.99511  & 0.99013   & 0.0000956092 \\
 0.5    &   0.986536    & 1.02809    & 3.97253  & 1.03108   & 0.000193669 \\
 0.6    &   0.941556    & 1.08059    & 3.92858  & 1.1269    & 0.0460541 \\
 0.7    &   0.871681    & 1.13783    & 3.8583   & 1.31712   & 0.179266 \\
 0.8    &   0.770921    & 1.19555    & 3.75332   & 1.54732  & 0.351768 \\
 0.9    &   0.625901    & 1.25148   & 3.59533   & 1.83464   & 0.583158 \\
 1.0     &   0.392439    & 1.30433   & 3.32443   & 2.23769   & 0.93336 \\
 1.01   &   0.35829     & 1.30942   & 3.28307   & 2.29239   & 0.982973 \\
 1.02   &   0.320004    & 1.31446   & 3.23615   & 2.35265   & 1.03819 \\
 1.03   &   0.275871    & 1.31946   & 3.18135   & 2.42081   & 1.10135 \\
 1.04   &   0.222438    & 1.32442   & 3.11393   & 2.5016    & 1.17718 \\
 1.05   &   0.15009     & 1.32934   & 3.02074   & 2.60817   & 1.27884 \\
1.05827 &    0.0        & 1.33333   & 2.82      & 2.82      & 1.48666\\
		\end{tabular}
	\end{ruledtabular}
\end{table*}
\endgroup
A numerical analysis of Eq.~(\ref{Delta2}) reveals that, for a fixed spin parameter $a$, the spacetime admits either two positive roots when $\ell < \ell_E$ or a single degenerate root when $\ell = \ell_E$. These cases correspond, respectively, to regular black hole solutions with distinct horizons and extremal black holes with coincident horizons~\cite{Kerr:1963ud,Chandrasekhar:1985kt}. The resulting parameter space is shown in Fig.~\ref{parspace}. For \textit{Model-III}, the light-blue region represents the set of parameters admitting black hole spacetimes, whereas the region beyond the red boundary curve corresponds to horizonless regular geometries.
Interestingly, the horizonless sector contains a distinct gray subregion in which closed photon-ring structures persist despite the absence of an event horizon. Such features are closely related to the existence of stable and unstable photon orbits and have been observed other ultracompact and horizonless objects~\cite{Cunha:2017qtt,Cunha:2018acu,Grenzebach:2014fha, Kumar:2020yem, Kumar:2020ltt}. Representative photon-ring configurations are displayed in Fig.~\ref{photonring}.
For $a=0.2M$, we find that a single unstable circular photon orbit exists throughout the black-hole regime, $0 \leq \ell \leq \ell_E = 1.044M$. In contrast, within the horizonless interval $\ell_E \leq \ell \leq \ell_P = 1.196M$, two circular photon orbits emerge: an inner stable orbit and an outer unstable orbit, as shown in Fig.~\ref{photonOrbit}. The appearance of multiple photon orbits is a characteristic feature of several regular and horizonless compact-object spacetimes and plays an important role in determining their lensing and shadow properties~\cite{Cardoso:2014sna,Cunha:2017qtt,Perlick:2021aok}.

A further noteworthy feature arises for extremal configurations. For small values of $\ell$ and the corresponding extremal spin $a=a_E$, the event-horizon radius coincides with the radius of the prograde circular photon orbit, similar to the behavior found in the extremal Kerr spacetime~\cite{Bardeen:1972fi,Teo:2003ltt}. As $\ell$ increases, however, this coincidence is lost and a finite separation,
\begin{equation}
\delta=\frac{r_m^{-}-r_E}{M}>0,
\end{equation}
develops between the two radii. This separation reflects the influence of the bounce parameter on the near-horizon geometry and provides a useful measure of the departure from the Kerr spacetime. Table~\ref{table1B} summarizes the extremal horizon radius together with the corresponding prograde and retrograde circular photon-orbit radii, $r^{-}$ and $r^{+}$, for representative extremal configurations of \textit{Model-III}.

\begin{figure*}
     \includegraphics[scale=0.65]{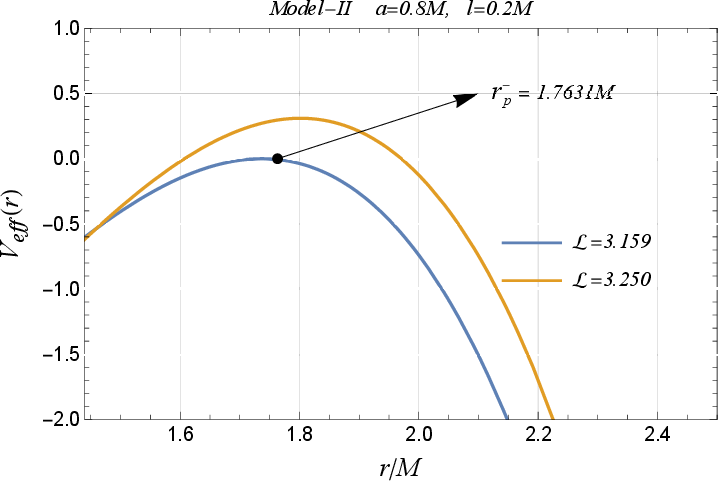}
     \hspace{1.5cm}
    \includegraphics[scale=0.65]{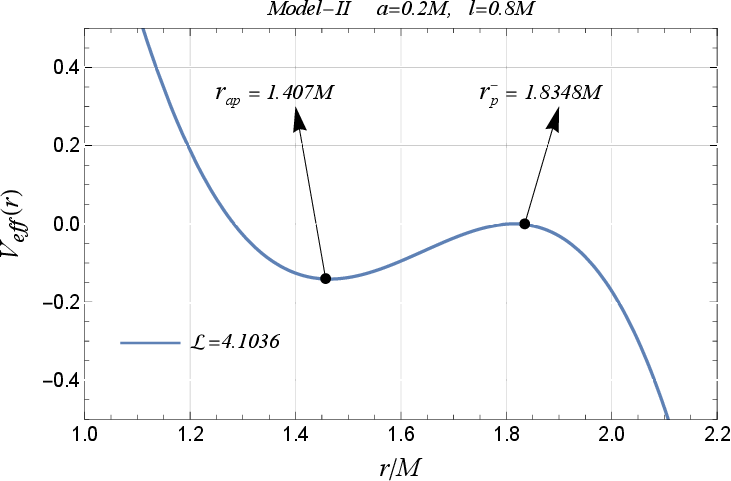}\\
    \includegraphics[scale=0.65]{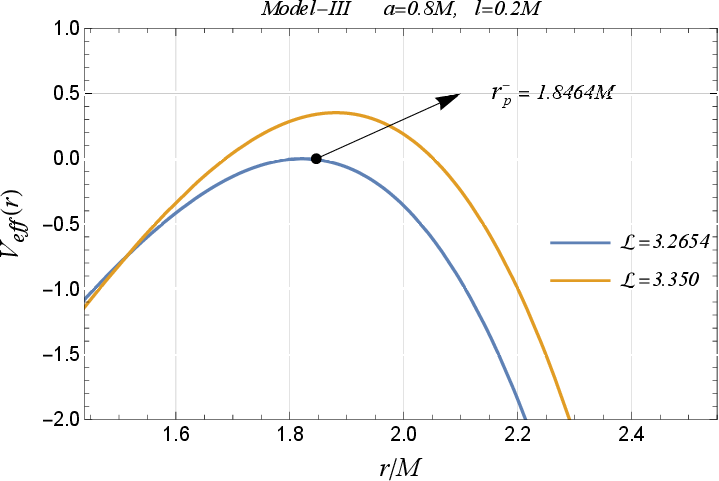}
     \hspace{1.5cm}
    \includegraphics[scale=0.65]{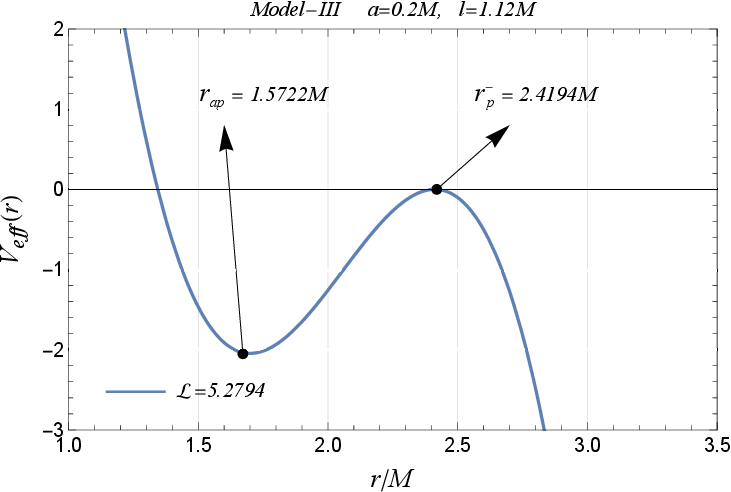}
    \caption{Effective potentials corresponding to light rays with impact parameters slightly larger and slightly smaller than that of the primary photon sphere for \textit{Model-II} (\textit{upper panels}) and \textit{Model-III} (\textit{lower panels}). For \textit{Model-II}, the choice $a=0.8M$ and $\ell=0.2M$ ($\ell<\ell_{E}=0.4405M$) yields a spacetime with only an unstable circular photon orbit located at $r_{m}=1.7631M$. On the other hand, for $a=0.2M$ and $\ell=0.8M$ ($\ell_{E}=0.7523M<\ell<\ell_{P}=0.81038M$), the spacetime is horizonless and admits both an unstable circular photon orbit at $r_{m}=1.8348M$ and a stable circular photon orbit at $r_{ap}=1.407M$. For \textit{Model-III}, the parameters $a=0.8M$ and $\ell=0.2M$ ($\ell<\ell_{E}=0.77456M$) correspond to a spacetime with only an unstable circular photon orbit at $r_{m}=1.84647M$. In contrast, for $a=0.2M$ and $\ell=1.12M$ ($\ell_{E}=1.0435M<\ell<\ell_{P}=1.196M$), both an unstable circular photon orbit at $r_{m}=2.41942M$ and a stable circular photon orbit at $r_{ap}=1.57229M$ are present. The emergence of a local minimum in the effective potential indicates the existence of stable photon orbits in the corresponding horizonless spacetimes.}
\label{Veff1}
\end{figure*}

\section{Strong gravitational lensing in black-bounce models}
\label{sec5}

Strong gravitational lensing is one of the most important observational consequences of GR and provides a powerful tool for probing the spacetime geometry in the vicinity of compact objects. In the strong-field regime, light rays can undergo large deflections and even execute multiple revolutions around the lens before reaching a distant observer, giving rise to relativistic images, photon rings, and black hole shadows~\cite{Darwin:1959zz,Luminet:1979nyg,Virbhadra:1999nm,Bozza:2002zj,Perlick:2004tq,Cunha:2018acu,Perlick:2021aok}. Such phenomena have attracted renewed interest following the horizon-scale EHT observations of supermassive black holes by the \cite{Akiyama:2019cqa,Akiyama:2019eap} and have been extensively investigated for black holes, wormholes, regular spacetimes, and other horizonless compact objects~\cite{Falcke:1999pj,Nedkova:2013msa,Tsukamoto:2016zdu,Cunha:2017qtt}.

In this section, we  wish to investigate the strong-field deflection of light in the rotating black-bounce spacetimes. For simplicity, we restrict photon motion to the equatorial plane ($\theta=\pi/2$), where the geodesic equations simplify considerably. We further assume that both the source and the observer are located in the asymptotically flat region of the spacetime, which is the standard setup adopted in strong gravitational lensing studies~\cite{Virbhadra:1999nm,Bozza:2002zj,Perlick:2004tq}. With these assumptions, the metric~(\ref{metric2}) reduces to
\begin{equation}\label{NSR}
ds^2=-A(r)\,dt^2+B(r)\,dr^2+C(r)\,d\phi^2-D(r)\,dt\,d\phi,
\end{equation}
where
\begin{eqnarray}\label{coeff}
A(r) &=& 1-\frac{2M(r)}{\Sigma(r)}, \qquad
B(r)=\frac{\Sigma(r)^2}{\Delta(r)}, \nonumber\\
C(r) &=& \frac{\left[\Sigma(r)^2+a^2\right]^2-a^2\Delta(r)}
{\Sigma(r)^2}, \qquad
D(r)=\frac{4aM(r)}{\Sigma(r)}.
\end{eqnarray}
The metric (\ref{NSR}) belongs to the general class of stationary and axisymmetric spacetimes commonly employed in studies of strong gravitational lensing and photon trajectories around rotating compact objects~\cite{Bozza:2002zj,Bozza:2003cp,Perlick:2004tq}.

Restricting the Hamilton-Jacobi equations to the equatorial plane ($\theta=\pi/2$), the radial equation of motion for null geodesics can be written as
\begin{eqnarray}\label{radialEq}
\left(\rho^2\frac{dr}{d\tau}\right)^2-V_{\rm eff}(r)=0,
\end{eqnarray}
where the effective radial potential is
\begin{equation}
V_{\rm eff}(r)
=
\left[(\Sigma^2+a^2)\mathcal{E}-aJ\right]^2
-\Delta(r)\left(a\mathcal{E}-J\right)^2 .
\end{equation}
The turning points of photon trajectories are determined by the condition $V_{\rm eff}(r)=0$, while circular photon orbits correspond to extrema of the effective potential~\cite{Chandrasekhar:1985kt,Teo:2003ltt,Cunha:2018acu}.
The metric functions $A(r)$, $B(r)$, $C(r)$, and $D(r)$ obey the conditions of asymptotic flatness in the limit $r \to \infty$.   Consequently, there exists an interval $(r_E,\infty)$ in which all metric functions remain regular and positive, where $r_E$ denotes the radius of the event horizon whenever it exists. Following the standard treatment of gravitational lensing in stationary axisymmetric spacetimes~\cite{Virbhadra:1999nm,Bozza:2002zj,Bozza:2003cp}, we introduce the impact parameter
\begin{equation}
u(r_0)=\frac{J}{\mathcal{E}},
\end{equation}
where $r_0$ is the radial turning point, i.e., the distance of closest approach of the photon trajectory. The critical value of the impact parameter is determined by the unstable circular photon orbit and plays a central role in the strong-deflection limit of gravitational lensing~\cite{Darwin:1959zz,Bozza:2002zj,Perlick:2021aok}. We set $\mathcal{E}=1$, so that the impact parameter $u$ overlaps with the conserved angular momentum $J$. From the null condition at the radial turning point $r_0$,  yields
\begin{equation}\label{impP}
u=J=
\frac{-D_0+\sqrt{D_0^2+4A_0C_0}}
{2A_0},
\end{equation}
where the subscript ``0'' indicates that the metric functions are evaluated at $r=r_0$~\cite{Bozza:2002zj,Bozza:2003cp}.
The black-bounce spacetimes listed in Table~\ref{table1} can admit multiple circular photon orbits. We denote the radius of the outermost circular null orbit by $r_m$. Depending on the behavior of the effective potential, these orbits may be either stable or unstable under radial perturbations~\cite{Chandrasekhar:1985kt,Cunha:2017qtt,Cunha:2018acu}. Light rays with the critical impact parameter $u(r_0)=u(r_m)$ asymptotically approach an unstable circular photon orbit, forming the photon sphere. The conditions determine the radius $r_m$
\begin{equation}
V_{\rm eff}(r_m)=V_{\rm eff}'(r_m)=0,
\end{equation}
together with
\begin{equation}
V_{\rm eff}''(r_m)<0,
\end{equation}
which ensures instability under radial perturbations.
In horizonless configurations, an additional circular null orbit may exist at a smaller radius $r_{ap}$. This orbit corresponds to a local minimum of the effective potential and is commonly referred to as an anti-photon sphere~\cite{Cunha:2017qtt,Cardoso:2014sna}. Its radius satisfies
\begin{equation}
V_{\rm eff}(r_{ap})=V_{\rm eff}'(r_{ap})=0,
\end{equation}
and
\begin{equation}
V_{\rm eff}''(r_{ap})>0.
\end{equation}
By construction, the anti-photon sphere lies inside the photon sphere, i.e., $r_{ap}<r_m$.
Both photon and anti-photon spheres exists in the horizonless compact objects, including wormholes and regular black-hole mimickers, and can lead to distinctive lensing  signatures~\cite{Cunha:2017qtt,Perlick:2021aok}. Moreover, asymptotic flatness implies that $V_{\rm eff}(r)\rightarrow -\infty$ as $r\rightarrow\infty$. Consequently, the outermost extremum of the effective potential must be a local maximum. Therefore, the outermost circular null orbit located at $r=r_m$ is necessarily unstable against outward radial perturbations, irrespective of the detailed form of the metric functions~\cite{Chandrasekhar:1985kt,Cunha:2018acu}.

We have plotted the $V_{\textit{eff}}(r)$ for \textit{Model-II} when $\ell<\ell_{E}$ and $\ell_{E}<\ell<\ell_{P}$ in Fig.~\ref{Veff1}. Particularly, we take $a=0.8M$,  $\ell=0.2M$($\ell<\ell_{E}=0.4405M$) and $a=0.2M$,  $\ell=0.8M$ (  $\ell_{E}=0.7523M<\ell<\ell_{P}=0.8104M$) such that in the former case only unstable circular orbits are formed at $r_{m}=1.7631M$. In contrast, in the latter case, the spacetime is horizonless and has both unstable and stable circular orbits at $r_{m}=1.835M$ and $r_{ap}=1.407M$, respectively. A similar analysis for \textit{Model-III}, reveals that for $a=0.8M$ and $\ell=0.2M$ ($\ell_{E}=0.7746M$) only unstable circular orbits are formed at $r_{m}=1.8465M$, while for $a=0.2M$,  $\ell=1.12M$ ($\ell_{E}=1.0435M<\ell<\ell_{P}=1.196M$) the unstable circular orbits are formed at $r_{m}=2.4194M$ and  stable circular orbits at $r_{ap}=1.5723M$.
\begingroup
\begin{table*}
	\caption{Summary of the strong gravitational lensing coefficients $\bar{a}$ and $\bar{b}$ for black hole-like spacetimes in \textit{Model-II} with parameters $a=0.8M$ and $\ell<\ell_E=0.44053$, and in \textit{Model-III} with parameters $a=0.8M$ and $\ell<\ell_E=0.77456$.}\label{table4A}
	\begin{ruledtabular}
		\begin{tabular}{c c c | c c c}
            \multicolumn{3}{c}{\textit{Model-II}}&
			\multicolumn{3}{c}{\textit{Model-III}}\\
   \cmidrule{1-3} \cmidrule{4-6} 
$\ell/M$ & $\bar{a} $ & $\bar{b} $ & $\ell/M$ & $\bar{a} $ & $\bar{b} $\\
\hline
 0.          & 1.916       & -1.587  & 0.            & 1.916    & -1.586 \\
 0.044053    & 1.923      & -1.599  & 0.07745       & 1.910     & -1.576 \\
 0.088106    & 1.943      & -1.635  & 0.15491       & 1.895     & -1.555 \\
 0.132159    & 1.980      & -1.702  & 0.23237       & 1.881     & -1.537 \\
 0.176212    & 2.034      & -1.806  & 0.30982       & 1.873     & -1.537 \\
 0.220265    & 2.119      & -1.965  & 0.38728       & 1.878     & -1.569 \\
 0.264318    & 2.244       & -2.211  & 0.46474       & 1.906    & -1.656 \\
 0.308372    & 2.440      & -2.621  & 0.54219       & 1.972     & -1.841 \\
 0.352425    & 2.787       & -3.400  & 0.61965       & 2.115    & -2.238 \\
 0.396478    & 3.601      & -5.465  & 0.69711       & 2.473     & -3.322 \\
		\end{tabular}
	\end{ruledtabular}
\end{table*}
\endgroup

\begingroup
\begin{table*}
	\caption{Summary of the strong gravitational lensing coefficients $\bar{a}$ and $\bar{b}$ for no-horizon spacetimes in \textit{Model-II} with parameters $a=0.2M$ and $\ell_E=0.752347<\ell<\ell_P=0.810381$, and in \textit{Model-III} with parameters $a=0.2M$ and $\ell_E=1.04353<\ell<\ell_P=1.19609$.}\label{table4B}
	\begin{ruledtabular}
		\begin{tabular}{c c c | c c c}
            \multicolumn{3}{c}{\textit{Model-II}}&
			\multicolumn{3}{c}{\textit{Model-III}}\\
   \cmidrule{1-3} \cmidrule{4-6} 
$\ell/M$ & $\bar{a} $ & $\bar{b} $ & $\ell/M$ & $\bar{a} $ & $\bar{b} $\\
\hline

 0.752347   & 1.703    & -1.410   & 1.04353    & 1.332     & -2.366 \\
 0.759871   & 1.758    & -1.552   & 1.0644     & 1.369     & -1.555 \\
 0.767394   & 1.825    & -1.737   & 1.08527    & 1.415     & -1.480 \\
 0.774918   & 1.909    & -1.986   & 1.10614    & 1.476     & -1.503 \\
 0.782441   & 2.020    & -2.342   & 1.12701    & 1.558     & -1.730 \\
 0.789965   & 2.178    & -2.902   & 1.14788    & 1.683     & -2.121 \\
 0.797488   & 2.436    & -3.942   & 1.16875    & 1.910     & -3.025 \\
 0.805012   & 3.023    & -6.842   & 1.18963    & 2.679     & -7.271 \\
		\end{tabular}
	\end{ruledtabular}
\end{table*}
\endgroup
\subsection{Deflection angle in the strong-field limit}
\label{sec6}

We now investigate the gravitational deflection of light in rotating black-bounce spacetimes in the strong-field regime. Particular attention is paid to photon trajectories that pass close to the photon sphere, where the bending angle becomes large and may even diverge logarithmically, giving rise to relativistic images and photon rings~\cite{Darwin:1959zz,Virbhadra:1999nm,Bozza:2002zj,Bozza:2003cp,Perlick:2021aok}. Our analysis applies to both black-hole and horizonless black-bounce configurations. The deflection angle can be obtained from the null geodesic equations (\ref{tuch})--(\ref{phiuch}), which completely determine photon trajectories in the equatorial plane. Following the strong-deflection formalism developed for stationary and axisymmetric spacetimes~\cite{Bozza:2002zj,Bozza:2003cp,Tsupko:2021yca}, the total bending angle associated with a photon trajectory having a closest-approach distance $r_0$ can be expressed as
\begin{equation}\label{bending2}
\alpha_D(r_0)=-\pi+I(r_0),
\end{equation}
where the integral term $I(r_0)$ accounts for the total change in the azimuthal coordinate along the photon path.

For the general stationary metric (\ref{NSR}), the integral takes the form
\begin{eqnarray}\label{EqA}
I(r_0)
&=&
2\int_{r_0}^{\infty}
P_1(r,r_0)\,P_2(r,r_0)\,dr,
\nonumber\\
P_1(r,r_0)
&=&
\frac{A_0\sqrt{B}\left(2AJ+D\right)}
{\sqrt{A_0C}\sqrt{4AC+D^2}},
\nonumber\\
P_2(r,r_0)
&=&
\frac{1}
{\sqrt{
A_0
-A\frac{C_0}{C}
+\frac{J}{C}
\left(A D_0-A_0D\right)
}}.
\end{eqnarray}
Here, the metric functions are evaluated along the photon trajectory, while quantities carrying the subscript ``0'' are evaluated at the turning point $r=r_0$. The integral (\ref{EqA}) remains finite for generic trajectories but develops a logarithmic divergence as $r_0$ approaches the radius of the unstable photon orbit $r_m$. This divergence is the hallmark of the strong-deflection regime and reflects the fact that photons can execute an arbitrarily large number of revolutions around the compact object before escaping to a distant observer~\cite{Darwin:1959zz,Bozza:2002zj,Bozza:2003cp}.
In the following, we analyze the behavior of Eq.~(\ref{EqA}) near the photon sphere and derive the corresponding strong-deflection coefficients for the black-bounce geometries under consideration. Particular emphasis will be placed on the differences between black-hole and horizonless configurations, especially in the presence of both photon and anti-photon spheres, which can lead to qualitatively distinct lensing signatures~\cite{Cunha:2017qtt,Perlick:2021aok,Tsukamoto:2020bjm}.
For convenience,  the metric functions evaluated at a general radial coordinate $r$ are denoted by $A(r)\equiv A$, $B(r)\equiv B$, $C(r)\equiv C$, and $D(r)\equiv D$. Quantities evaluated at the closest-approach distance $r_0$ are indicated by a subscript ``0'', namely $A(r_0)\equiv A_0$, $B(r_0)\equiv B_0$, and similarly for the other metric functions.

The quantity $I(r_0)$ measures the total azimuthal shift accumulated by a photon as it propagates through the spacetime. In the absence of gravity, the photon travels along a straight line and one simply has $I(r_0)=\pi$. Gravitational bending caused by a compact object increases the path length in the angular direction, leading to $I(r_0)>\pi$.
 As the distance of closest approach decreases and approaches the radius of the unstable photon orbit $r_m$, the deflection becomes increasingly strong. In the limit $r_0\rightarrow r_m$, the photon can orbit the compact object an arbitrarily large number of times before escaping, and the integral $I(r_0)$ diverges logarithmically~\cite{Darwin:1959zz,Bozza:2002zj,Bozza:2003cp,Perlick:2021aok}. This logarithmic divergence forms the basis of the strong-deflection limit (SDL) formalism.

In general, the integral (\ref{EqA}) cannot be evaluated analytically in closed form. Nevertheless, its behavior near the photon sphere can be extracted by isolating the divergent contribution and performing an appropriate expansion around $r_0=r_m$~\cite{Bozza:2002zj,Bozza:2003cp,Tsukamoto:2016jzh,Nascimento:2020ime}.

\subsection{Case-I: \texorpdfstring{$0 \leq \ell \leq \ell_E$}{0 <= l <= l\_E}}

In this parameter range, the spacetime describes a black hole with an event horizon and a single unstable photon sphere located outside the horizon. To obtain the deflection angle in the strong-deflection limit, we analyze the behavior of the integral in Eq.~(\ref{EqA}) as the closest-approach distance $r_0$ approaches the photon-sphere radius $r_m$. Following the standard strong-deflection formalism developed by Bozza and subsequently extended to rotating and horizonless spacetimes~\cite{Bozza:2002zj,Bozza:2003cp,Tsukamoto:2016jzh,Nascimento:2020ime}, we introduce the variable
\begin{equation}
z=1-\frac{r_0}{r},
\end{equation}
which transforms the integration domain $r\in[r_0,\infty)$ into the finite interval $z\in[0,1)$. This change of variables makes the divergent behavior of the integrand near the photon sphere explicit and facilitates the extraction of the logarithmic contribution to the deflection angle that characterizes the strong-deflection regime~\cite{Darwin:1959zz,Bozza:2002zj}.

This method captures the behavior of photons near the photon sphere and yields an analytical expression for the deflection angle \cite{Bozza:2002zj} as follows:
\begin{eqnarray}\label{def4}
\alpha_{D}(u) &=& \bar{a} \log\left(\frac{u}{u_{m}} - 1\right) + \bar{b} + \mathcal{O}(u - u_{ps}), \,
\end{eqnarray}
with $\bar{a}$ and $\bar{b}$, being the lensing coefficients and are tabulated for our black bounce models in Table \ref{table4A}.

\subsection{Case II: $\ell_{E}<\ell<\ell_{P}$}
For $\ell_{E}<\ell<\ell_{P}$, the spacetime has a photon sphere, 
and an antiphoton sphere inside it. The infinite sets  of images are formed as discussed accordingly as below. 

\subsubsection{Images slightly outside of the primary photon sphere~(ISOP)}
We can use the formulas in Sec.~\ref{sec6}.
Parameters $\bar{a}$ and $\bar{b}$ in the deflection angle~(\ref{def4}) for the images slightly outside of the primary photon sphere  in the strong deflection limit are tabled in Table \ref{table4B}.

\subsubsection{Images slightly inside the primary photon sphere (ISIP)}

We first consider the class of photon trajectories that produce images slightly inside the primary photon sphere. In this case, the photons penetrate the photon sphere, traverse the antiphoton sphere, and are subsequently reflected by the effective potential before escaping to a distant observer. As a result, the corresponding images appear slightly inside the critical curve associated with the primary photon sphere. The qualitative behavior of the effective potential governing such trajectories is shown in Fig.~\ref{Veff1}. For trajectories that approach the photon sphere from  the interior (i.e., from smaller radii), the strong deflection limit 
analysis remains valid under the assumption that the radial potential near 
the photon sphere can be approximated by a quadratic expansion, as is 
standard in the Bozza formalism \cite{Bozza:2002zj}. This expansion is 
justified provided that the impact parameter is sufficiently close to its 
critical value and that the photon sphere is nondegenerate. For the 
interior trajectories considered here, the same logarithmic divergence of 
the deflection angle arises, with the divergence coefficient determined by 
the second derivative of the effective potential evaluated at the photon 
sphere radius. However, the regular part of the deflection integral may 
differ for interior and exterior trajectories due to the different 
integration domains.

For photons with impact parameters slightly smaller than the critical value $u_m$, the radial motion extends beyond the photon sphere and reaches a turning point located inside the antiphoton sphere. After reflection, the photons cross the antiphoton sphere and photon sphere once again before reaching the observer. Following Ref.~\cite{Shaikh:2019itn}, we introduce the variable
\begin{equation}\label{eq34}
z=1-\frac{r_m}{r},
\end{equation}
which is particularly convenient for analyzing the deflection angle in the vicinity of the photon sphere.

The smaller positive root of the effective potential corresponding to the critical impact parameter $u=u_m$ is denoted by $r=r_c$. By construction,
\begin{equation}
u_c \equiv u(r_c)=u_m.
\end{equation}
The radius $r_c$ marks the inner turning point of the critical trajectory and plays a central role in the formation of images inside the primary photon sphere. Here and throughout the following discussion, quantities carrying the subscript ``c'' are evaluated at $r=r_c$. Similar image structures associated with photon and antiphoton spheres have been reported in several horizonless compact-object spacetimes, where the presence of stable and unstable circular photon orbits leads to rich lensing phenomenology~\cite{Shaikh:2019itn,Cunha:2017qtt,Perlick:2021aok}.
 By using $z$ in Eq.~(\ref{eq34}), we rewrite Eq.~(\ref{EqA}) as
\begin{eqnarray}
I(r_0,r_m)=\int_{1-r_m/r_0}^{1} R(z,r_0,r_m)f(z,r_0,r_m)~dz,
\end{eqnarray}
where 
\begin{eqnarray}
R(z,r_0,r_m) = \frac{2 r_m P_1(z,r_0,r_m)}{(1-z)^2},\nonumber\\
f(z,r_0,r_m)=P_2(z,r_0,r_m)=\frac{1}{\sqrt{G(z,r_0,r_m)}}
\end{eqnarray}
with $r = r_m/(1-z)$. Further, we expand $G(z,r_0,r_m)$ around $r=r_m$ or $z=0$ as
\begin{equation}
G(z,r_0) = \alpha+\beta~z+\eta~z^2+O(z^3),
\end{equation}
where $\alpha(r_0,r_m)$, $\beta(r_0,r_m)$, and $\gamma(r_0,r_m)$ are given by
\begin{eqnarray}
\alpha &&= \frac{1}{C_m}\Big(A_0 C_m - A_m C_0 + J\,(A_m D_0 - A_0 D_m)\Big) \nonumber\\
\beta &&= \frac{r_m}{C_m^2}
\Bigg\{ C_0(A_m C'_m - C_m A'_m )  + J\big(D_0(C_m A'_m \nonumber\\
&& - A_m C'_m) + A_0(D_m C'_m - C_m D'_m)\big)
\Bigg\}\nonumber \\
\eta && =  \frac{r_m}{2 C_m^3} \Bigg\{
(C_0 - J D_0)\,\Big(
2 C_m (C_m - r_m C'_m)\, A'_m \nonumber\\
&&
+ r_m C_m^2\, A''_m  - 2 C_m A_m C'_m 
+ 2 r_m A_m (C'_m)^2
\Big)  \nonumber\\
&&+ J A_0 \,\Big(
2 C_m D_m C'_m 
- 2 r_m D_m (C'_m)^2 
+ 2 C_m \nonumber\\
&& (r_m C'_m  - C_m)\, D'_m 
- r_m C_m^2\, D''_m
\Big) + r_m C_m 
\nonumber\\
&&
\Big(- A_m (C_0   - J D_0) + J A_0 D_m
\Big)\, C''_m
\Bigg\}.
\end{eqnarray}
Here functions are represented such that  $A(r_m)\equiv A_m$ and $A(r_c)\equiv A_c$ and so on. It must be noted that the height of effective potential at $r=r_m$ and $r=r_c$ is the same, thus when impact parameter approaches the critical value $J_m$ the turning point approaches the radius $r_c$, thus  $J_0\to J_m$ as $r_0\to r_c$. Moreover, using the equations 
\begin{eqnarray}
   A(r)J(r)^2+D(r)J(r)-C(r)=0\Bigg|_{r=r_m,~r_0}
\end{eqnarray}
\begin{equation}
    (A_mC'_m-A'_mC_m)+J_m(A'_mD_m-A_mD'_m)=0,
\end{equation}
in the strong deflection limit $r_0 \rightarrow r_c$ or $J \rightarrow J_c=J_m$, we have,
\begin{eqnarray}
\alpha_c &\rightarrow& +0, \\
\beta_c  &\rightarrow& -0, \\
\eta_c &\rightarrow& \frac{r_m^2 A_c J_m}{2 C_m}
\Big(J_m A''_m - D''_m \nonumber\\
&& + \frac{C''_m}{C_m}(D_m - A_m J_m)\Big).
\end{eqnarray}
Hence, we have 
\begin{equation}
   G_m(z,r_c)=\eta_c z^2 + \mathcal{O}(z^3).
\end{equation}
The term $I(r_0)$, thus,  diverges logarithmically in the strong deflection limit $r_0 \rightarrow r_c$.  To extract the logarithmic divergence we split the integral $I(r_0)$ into divergent part $I_D(r_0)$ and a regular part $I_R(r_0)$ such that
\begin{equation}
    I(r_0,r_m) = I_R(r_0,r_m) + I_D(r_0,r_m)
\end{equation}
where
\begin{eqnarray}
   I_R(r_0,r_m) &=& \int_{1-r_m/r_0}^{1} \Big(R(z,r_0,r_m)f(z,r_0,r_m)\nonumber\\
   &-& R(0,r_0,r_m)f_D(z,r_0,r_m)\big)dz,\nonumber\\
 I_D(r_0,r_m) &=& \int_{1-r_m/r_0}^{1} R(0,r_0,r_m)f_D(z,r_0,r_m)dz,\nonumber\\
 f_D(z,r_0,r_m) &=& \frac{1}{\sqrt{\alpha + \beta~z + \eta~z^2 }}\quad\quad
\end{eqnarray}    
The integral $ I_D(r_0,r_m)$ can be solved to 
\begin{widetext}
\begin{eqnarray}\label{DivTerm}
&& I_D(r_0,r_m) = 
\frac{R(0,x_0,x_m)}{\sqrt{\eta}} ~\log \frac{\beta + 2\eta  +  2\sqrt{\eta(\alpha + \beta + \eta)}}{\beta + 2\eta\Big(1 - \frac{r_m}{r_0}\Big)   + 2\sqrt{\eta\left(\alpha + \beta\left(1 - \frac{r_m}{r_0}\right) + \eta\left(1 - \frac{r_m}{r_0}\right)^2\right)} \quad }
\end{eqnarray}    
\end{widetext}

Using $\delta = 1 - \frac{r_m}{r_0}$, the numerator $(N)$ and denominator $(D)$ of the Logarithmic term in Eq.~(\ref{DivTerm}) under the strong lensing condition wherein the terms $\alpha$ and $\beta$ can be treated as small parameters  can be simplified as follows
\begin{eqnarray}
\frac{N}{D} &=& \frac{\beta + 2\eta + 2\sqrt{\eta}\sqrt{\alpha+\beta+\eta} }{\beta + 2\eta \delta + 2\sqrt{\eta}\sqrt{\alpha + \beta \delta + \eta \delta^2}}\nonumber\\
 &\approx& \frac{\beta + 2\eta + 2\eta \left(1 + \frac{\alpha+\beta}{2\eta} + \cdots \right)}{\beta + 2\eta \delta + 2\sqrt{\eta}\sqrt{\alpha}}\nonumber\\
 &\approx& \frac{-4\eta_c\delta_c}{\alpha_c}
\end{eqnarray}
Thus 
\begin{equation}\label{eqnId}
I_D(r_0) = \frac{R(0,x_0,x_m)}{\sqrt{\eta_c}} \log\left(
\frac{-4\eta_c\left(1-\frac{r_m}{r_c}\right)}{\alpha_c}
\right).
\end{equation}
We can also write the expansion of $\alpha$ about $r_c$ by defining $\varepsilon=r_0-r_c$ as follows
\begin{eqnarray}
 \alpha &=&  \frac{1}{C_m}\Big(C_m(A_c+A'_c\varepsilon+\cdots) - A_m (C_c+C'_c\varepsilon+\cdots) \nonumber\\ 
 &+& (J_c+J'_c\varepsilon+\cdots)(A_m (D_c+D'_c\varepsilon+\cdots) 
 \nonumber\\ &-& D_m(A_c+A'_c\varepsilon+\cdots))  
\end{eqnarray}

\begin{widetext}
\begin{eqnarray}\label{eqnalpha}
\alpha= \frac{1}{C_m} \Big(  \underbrace{(A_cC_m-A_mC_c)+J_c(A_mD_c-A_cD_m)}_{\text{Constant Part}}  +  \underbrace{(A'_cC_m-A_mC'_c)+J_c(A_mD'_c-A_cD'_m)+J'_c(A_mD_c-D_mA_c)}_{\text{Linear Part}}\varepsilon \Big) \quad\quad  
\end{eqnarray}
\end{widetext}
Using the  strong deflection limit $r_0 \rightarrow r_c$ or $J \rightarrow J_c=J_m$, the linear part  vanishes and we are left with
\begin{eqnarray}
  \alpha_c \approx \frac{\mathscr{L}}{C_m}(r_0 - r_c).
\end{eqnarray}
where $\mathcal{L}$ is the linear part of the Eq.~(\ref{eqnalpha}). 
Expanding $J$ around $r_c$ and using $J_c=J_m$, we have 
\begin{equation}
    J-J_m=J'_c(r_0-r_c)
\end{equation}
Thus 
\begin{eqnarray}
    \alpha_c = \frac{-\mathscr{L}J_m}{C_mJ'_c}\Big(1-\frac{J}{J_m}\Big)
\end{eqnarray}
Hence, $I_D(x_0)$, simplifies to 
\begin{equation}
    I_D(r_0) = \frac{R(0,x_0,x_m)}{\sqrt{\eta_c}}~ \log\left(
\frac{4\eta_c\left(\frac{r_m}{r_c}-1\right)
}{ \frac{-\mathscr{L}J_m}{C_mJ'_c}\Big(1-\frac{J}{J_m}\Big)}
\right).
\end{equation}

Finally,  we obtain
\begin{align}\label{divterm2}
I_D(r_0)
&=
-\frac{R(0,x_c,x_m)}{\sqrt{\eta_c}} \log\left(1-\frac{J}{J_m}\right)\nonumber \\
&\quad +
\frac{R(0,x_c,x_m)}{\sqrt{\eta_c}}
\log\left[
\frac{-4\eta_c C_m J'_c}{\mathscr{L}J_m}
\left(\frac{r_m}{r_c}-1\right)
\right]
\end{align}
The regular part $I_{R}$ can be expanded in the powers of $r_0-r_c$ and taking only the leading part which can be integrated numerically as follows
\begin{eqnarray}
I_R(r_c,r_m) &=& \int_{1-r_m/r_c}^{1} f_R(z,r_c,r_m)dz
\end{eqnarray}
where $f_R(z,r_c,r_m)$ is given by
\begin{eqnarray}
 f_R(z,r_c,r_m) &=& \Big(R(z,r_c,r_m)f(z,r_c,r_m)\nonumber\\
   &-& R(0,r_c,r_m)f_D(z,r_c,r_m)\big)dz,
\end{eqnarray}
Finally, the deflection angle in the strong deflection limit can be written as 
\begin{eqnarray}\label{defInside}
    \alpha_{D}(u) &=& -\bar{c}\log\left(1-\frac{u}{u_m} \right) + \bar{d}.
\end{eqnarray}
where 
\begin{eqnarray}\label{deflection2}
    \bar{c} &=& \frac{R(0,x_c,x_m)}{\sqrt{\eta_c}}\nonumber\\
    \bar{d} &=& \bar{c}~ 
\log\left(
\frac{-4\eta_c C_m J'_c}{\mathscr{L}J_m}
\left(\frac{r_m}{r_c}-1\right)
\right) \nonumber\\ &+& I_R(r_c,r_m) -\pi
\end{eqnarray}
Parameters $\bar{c}$ and $\bar{d}$ in the deflection angle~(\ref{deflection2}) for the images slightly inside the primary photon sphere 
in the strong deflection limit~$b \rightarrow b_{\mathrm{m}}-0$ are shown in Table \ref{table4B}.

\begingroup
\begin{table*}
	\caption{Summary of the strong gravitational lensing coefficients $\bar{c}$ and $\bar{d}$ for no-horizon spacetimes in \textit{Model-II} with parameters $a=0.1M$ and $\ell_E=0.752347<\ell<\ell_P=0.810381$, and in \textit{Model-III} with parameters $a=0.1M$ and $\ell_E=1.04353<\ell<\ell_P=1.19609$.}\label{table4B3}
	\begin{ruledtabular}
		\begin{tabular}{c c c | c c c}
            \multicolumn{3}{c}{\textit{Model-II}}&
			\multicolumn{3}{c}{\textit{Model-III}}\\
   \cmidrule{1-3} \cmidrule{4-6} 
$\ell/M$ & $\bar{c} $ & $\bar{d} $ & $\ell/M$ & $\bar{c} $ & $\bar{d} $\\
\hline
0.765462 & 3.080  & 5.252    & 1.05546   & 2.492   & 5.025 \\
0.771077 & 3.132  & 5.109    & 1.07765   & 2.550   & 4.788 \\
0.776692 & 3.189  & 4.943    & 1.09846  & 2.620   & 4.505 \\
0.782307 & 3.252  & 4.748    & 1.12038  & 2.707    & 4.153 \\
0.787922 & 3.323   & 4.521   & 1.14230    & 2.819   & 3.695 \\
0.793536 & 3.405  & 4.253    & 1.16423  & 2.970    & 3.058 \\
0.799151 & 3.497 & 3.932     & 1.18615   & 3.189   & 2.077 \\
0.804766 & 3.605 & 3.544     & 1.20808   & 3.555   & 0.281 \\
0.810381 & 3.733  & 3.061    & 1.230      & 4.400   & -4.580 \\
		\end{tabular}
	\end{ruledtabular}
\end{table*}
\endgroup

\begingroup
\begin{table*}
	\caption{Summary of the strong gravitational lensing coefficients $\bar{c}$ and $\bar{d}$ for no-horizon spacetimes in \textit{Model-II} with parameters $a=0.2M$ and $\ell_E=0.752347<\ell<\ell_P=0.810381$, and in \textit{Model-III} with parameters $a=0.2M$ and $\ell_E=1.04353<\ell<\ell_P=1.19609$.}\label{table4B4}
	\begin{ruledtabular}
		\begin{tabular}{c c c | c c c}
            \multicolumn{3}{c}{\textit{Model-II}}&
			\multicolumn{3}{c}{\textit{Model-III}}\\
   \cmidrule{1-3} \cmidrule{4-6} 
$\ell/M$ & $\bar{c} $ & $\bar{d} $ & $\ell/M$ & $\bar{c} $ & $\bar{d} $\\
\hline
 0.77       & 4.501   & 23.790   & 1.04353    & 2.785   & 8.385 \\
 0.774286   & 4.716   & 19.416   & 1.06440    & 2.885   & 7.437 \\
 0.778571   & 4.984   & 17.281   & 1.08527    & 3.013   & 6.664 \\
 0.782857   & 5.332   & 15.992   & 1.10614    & 3.186   & 5.989 \\
 0.787143   & 5.809   & 15.259   & 1.12701    & 3.437   & 5.362 \\
 0.791429   & 6.523   & 15.126   & 1.14788    & 3.849   & 4.736 \\
 0.795714   & 7.768   & 16.134   & 1.16875    & 4.740   & 4.025 \\
 0.8        & 10.869   & 21.641  & 1.18963    & 13.690   & 4.941 \\
		\end{tabular}
	\end{ruledtabular}
\end{table*}
\endgroup

\begin{table*}
\setlength{\tabcolsep}{8pt}
\caption{Observables in the strong deflection limits for given $a/M$ and $\ell/M$ for the case of Sgr A* considered as \textit{Model-II} with  $D_{OS}=16$~KPc, $D_{OL}=D_{LS}=8$~KPc and  $M\equiv 4\times 10^6 M_{\odot}$.
 The diameter of the brightest image~$2\theta_{1}$, the diameter of the photon sphere~$2\theta_{\infty}$, 
 and angular diameter of images inside the photon sphere $2\theta_{-n}$,
 the difference of the radii of the brightest image and the photon sphere ${s_{\pm 1}}=\theta_{\pm 1}-\theta_\infty$, 
 the magnification of the  brightest image $\mu_{1} $ outside the photon sphere and those of images  inside   the photon sphere  $\mu_{-n}$ for the source angle $\beta=1$ arcsecond, 
the ratio of the magnification of the brightest image to the other images $\mathcal{R}_{\pm 1} = \mu_1/\sum^\infty_{n=2} \mu_{\pm n}$
 are shown.  
}\label{Table8A}
\begin{center}
\begin{tabular}{c|c| c c c c c c c c c}
\hline
$a$\Big/ $\ell$ &   &$0$ & $0.2M$ & $0.4M$ & $0.5M$ & $0.75M$ &  $0.77M$ &   $0.8M$  & $0.82M$  &  $0.84M$  \\ 
\hline

\multirow{13}{*}{$0.1M$}
& $2\theta_{1}$~[$\mu$as] ~        &  49.73 & 49.49  & 48.74 & 48.134 & 45.62 & 45.32 & 44.82 & 44.17 & 43.1   \\

& $2\theta_{\infty}$~[$\mu$as]     & 49.5 & 49.25 & 48.45  & 47.79 & 44.93 & 44.57 & 43.94 & 43.16 & 42.48  \\

& $2\theta_{-3}$~[$\mu$as]          & ... & ... & ... &...  &...& 44.29 &	43.63 &	43.12 &	42.59 \\
& $2\theta_{-2}$~[$\mu$as]          & ... & ... & ... & ...&... & 42.50  & 42.09 & 41.92 & 42.03 \\
& $2\theta_{-1}$~[$\mu$as]          & ... & ... & ... & ...&... & 29.12 & 32.86 & 36.15 & 40.25 \\

& $\mu_{1}\times 10^{17}$ ~       &0.23 & 0.24 & 0.278 & 0.314 & 0.55 & 0.59 & 0.65 &0.67 & 0.32\\

& $\mu_{-3}\times 10^{17}$~                       & ... & ... & ... & ... &... & 0.948 &  0.932 & 0.833 &  0.492 \\
& $\mu_{-2}\times 10^{17}$~                       & ... &...  & ... & ... & ... &  6.81   &  5.379 & 3.863  &  1.532  \\
& $\mu_{-1}\times 10^{17}$~                       & ...  & ... & ... & ... & ... & 34.93  &  25.13 &  15.89 &  4.634 \\

& $s_{1}$~~[$\mu$as]                & 0.115 & 0.12 & 0.145 & 0.172 & 0.345 & 0.375 &  0.44 & 0.505 & 0.31  \\
& $s_{-1}$~~[$\mu$as]               & ... & ... & ... &... &...  & -7.721 & -5.539  & -3.647 & -1.298 \\

& $\mathcal{R}_{1}$ ~               & 416.5 & 389. & 309.7 & 253.5 & 82.05 & 67.46 & 45.55 & 23.31 & 6.563  \\
& $\mathcal{R}_{-1}$~               & ... & ... & ... & ... & ... &  4.418 & 3.866 & 3.231 & 2.056 \\
\hline

\multirow{13}{*}{$0.2M$}
& $2\theta_{1}$~~[$\mu$as]           & 47.7 & 47.45 & 46.63 & 45.96 &  43.1 & 42.71 & 41.75 & ... &...\\
& $2\theta_{\infty}$~[$\mu$as]       & 47.42 & 47.14 & 46.26 & 45.53 & 42.11 & 41.62 & 40.68 & ... &...\\

& $2\theta_{-3}$~[$\mu$as]            & ... &...  & ... & ... & ... & 41.51	& 40.42 &... &...\\
& $2\theta_{-2}$~[$\mu$as]            & ... & ... & ... & ... & ... & 41.16 & 40.21  & ... & ...\\ 
& $2\theta_{-1}$~[$\mu$as]            & ... & ... & ... & ... & ...  & 39.78 &	39.84 & ... & ... \\

& $\mu_{1} \times 10^{17}$ ~         & 0.27 & 0.28 & 0.32 & 0.37 & 0.69 & 0.72 & 0.58 & ... &...  \\

& $\mu_{-3} \times 10^{17}$~                        & ... & ... & ... & ... & ... &  0.252  &	0.240  & ... & ... \\
& $\mu_{-2} \times 10^{17}$~                        & ... & ... & ... & ... & ... &  1.01	&   0.425 & ... &...  \\
& $\mu_{-1} \times 10^{17}$~                        & ... & ... & ... & ... & ... & 3.941   &	0.750  & ... &...  \\

& $s_{1}$~~[$\mu$as]                 & 0.14 & 0.155 & 0.185 & 0.215 & 0.495 & 0.545 & 0.535 &... &...\\
& $s_{-1}$~~[$\mu$as]                & ... &...  & ... & ... & ... &-0.92 & -0.42 &...&... \\

& $\mathcal{R}_{1}$ ~                & 317.9 & 294.6 & 227.7 & 180.8 & 40.85 & 29.12 & 10.69 & ... &...   \\
& $\mathcal{R}_{-1}$~                & ... & ... & ... & ... & ... &2.93 & 0.77 &...  \\
\hline

\multirow{5}{*}{$0.8M$}
& $\theta_{1}$~~[$\mu$as]           & 33.59 & 33.04 & 30.51  &...& ...& ...  &... &... & ... \\
& $\theta_{\infty}$~[$\mu$as]       & 32.09 & 31.32 & 27.75  &...& ...& ...  &... &... & ... \\
& $\mu_{1} \times 10^{17}$ ~        &  6.34 &  6.936 & 8.4 &...& ...& ...  &... &... & ... \\
& $s_{1}$ ~~[$\mu$as]               & 0.75 & 0.86 & 1.38 &...& ...& ...  &... &... & ... \\
& $\mathcal{R}_{1}$ ~               & 26.13 & 20.13 & 4.649 &...& ...& ...  &... &... & ... \\
\hline
\hline
\end{tabular}
\end{center}
\end{table*}

\begin{table*}
\setlength{\tabcolsep}{10pt}
\caption{Observables in the strong deflection limits for given $a/M$ and $\ell/M$ for the case of Sgr A* considered as \textit{Model-III} with  $D_{OS}=16$~KPc, $D_{OL}=D_{LS}=8$~KPc and  $M\equiv 4\times 10^6 M_{\odot}$.
 The diameter of the brightest image~$2\theta_{1}$, the diameter of the photon sphere~$2\theta_{\infty}$, 
 and angular diameter of images inside the photon sphere $2\theta_{-n}$,
 the difference of the radii of the brightest image and the photon sphere ${s_{\pm 1}}=\theta_{\pm 1}-\theta_\infty$, 
 the magnification of the  brightest image $\mu_{1} $ outside the photon sphere and those of images  inside   the photon sphere  $\mu_{-n}$ for the source angle $\beta=1$ arcsecond, 
the ratio of the magnification of the brightest image to the other images $\mathcal{R}_{\pm 1} = \mu_1/\sum^\infty_{n=2} \mu_{\pm n}$
 are shown.  
}\label{Table8B}
\begin{center}
\begin{tabular}{c| c |c c c c c c c c}
\hline
$a$\Big/ $\ell$ &   &$0$ & $0.25M$ & $0.5M$ &  $0.75M$ & $0.9M$ &  $1.056M$ &   $1.1M$  & $1.2M$   \\ 
\hline

\multirow{13}{*}{$0.1M$}
& $2\theta_{1}$~[$\mu$as] ~        & 49.73 & 49.89 & 50.24 & 50.57 & 50.67 & 50.59 & 50.52 & 51.89  \\
& $2\theta_{\infty}$~[$\mu$as]     & 49.5 & 49.66 & 50. & 50.3 & 50.35 & 50.18 & 50.05 & 51.24 \\

& $2\theta_{-3}$~[$\mu$as]          & ... & ... & ... &...  &...& 50.079 & 49.948 & 51.11 \\
& $2\theta_{-2}$~[$\mu$as]          & ... & ... & ... & ...&... &  48.961 & 48.901 & 50.37 \\
& $2\theta_{-1}$~[$\mu$as]          & ... & ... & ... & ...&... & 35.106 & 37.447 & 45.685 \\

& $\mu_{1}\times 10^{17}$ ~       & 0.9232 & 0.9225 & 0.9501 & 1.063 & 1.223 & 1.565 & 1.722 & 2.245\\

& $\mu_{-3}\times 10^{17}$~                       & ... & ... & ... & ... &... &  0.4768 & 0.486 & 0.5001 \\
& $\mu_{-2}\times 10^{17}$~                       & ... &...  & ... & ... & ... &   5.7804 & 5.199 & 3.1401 \\
& $\mu_{-1}\times 10^{17}$~                       & ...  & ... & ... & ... & ... & 51.393 & 43.578 & 18.144 \\

& $s_{1}$~~[$\mu$as]                & 0.115 & 0.115 & 0.12 & 0.135 & 0.16 & 0.205 & 0.235 & 0.325 \\
& $s_{-1}$~~[$\mu$as]               & ... & ... & ... &... &...  &-7.536 & -6.30 & -2.78  \\

& $\mathcal{R}_{1}$ ~               & 416.5 & 418.8 & 404.2 &  341.3 & 269.8 & 166.4 & 132.4 & 49.99  \\

& $\mathcal{R}_{-1}$~               & ... & ... & ... & ... & ... &  8.159 & 7.6 & 4.86\\
\hline

\multirow{13}{*}{$0.2M$}
& $2\theta_{1}$~~[$\mu$as]           & 47.70 & 47.87 & 48.24 & 48.58 & 48.65 & 48.52 & 48.40 & ... \\

& $2\theta_{\infty}$~[$\mu$as]       & 47.42 & 47.59 & 47.95 & 48.25 & 48.26 & 47.97 & 47.77 & ... \\

& $2\theta_{-3}$~[$\mu$as]            & ... &...  & ... & ... & ... & 47.831 & 47.61 &...\\
& $2\theta_{-2}$~[$\mu$as]            & ... & ... & ... & ... & ... &46.735 & 46.534  & ... \\ 
& $2\theta_{-1}$~[$\mu$as]            & ... & ... & ... & ... & ...  & 36.736 & 38.533  & ...\\

& $\mu_{1} \times 10^{17}$ ~         & 1.065 & 1.063 & 1.095 & 1.237 & 1.446 & 1.906 & 2.112 & ...  \\

& $\mu_{-3} \times 10^{17}$~                        & ... & ... & ... & ... & ... &  0.5501 & 0.614 & ...  \\
& $\mu_{-2} \times 10^{17}$~                        & ... & ... & ... & ... & ... &  4.906 & 4.473 & ...   \\
& $\mu_{-1} \times 10^{17}$~                        & ... & ... & ... & ... & ... & 35.192 & 27.59 & ...   \\

& $s_{1}$~~[$\mu$as]                 & 0.14 & 0.14 & 0.145 & 0.165 & 0.195 & 0.275 & 0.315 &... \\
& $s_{-1}$~~[$\mu$as]                & ... &...  & ... & ... & ... &-5.615 & -4.621 &... \\

& $\mathcal{R}_{1}$ ~                & 3317.9 & 320.4 & 308.9 & 254.9 & 192.8 & 103.3 & 74.25 & ...   \\
& $\mathcal{R}_{-1}$~                & ... & ... & ... & ... & ... &6.372 & 5.323&...  \\
\hline

\multirow{5}{*}{$0.95M$}
& $\theta_{1}$~~[$\mu$as]           & 28.48 & 29.12 & 29.49  &...& ...& ...  &... &...  \\
& $\theta_{\infty}$~[$\mu$as]       & 25.6 & 26.4 & 26.82  &...& ...& ...  &... &...  \\
& $\mu_{1} \times 10^{17}$ ~        & 3.472 &  3.575 & 3.189&...& ...& ...  &... &...  \\
& $s_{1}$ ~~[$\mu$as]               & 1.44 & 1.36 & 1.335 &...& ...& ...  &... &...\\
& $\mathcal{R}_{1}$ ~               & 5.263 & 6.606 & 4.46 &...& ...& ...  &... &...  \\
\hline
\hline
\end{tabular}
\end{center}
\end{table*}

\section{Observables in the strong deflection limit}
\label{sec7}
Having derived the deflection angle in the strong-deflection limit for photon trajectories both outside and inside the photon sphere, we now turn to the corresponding lensing observables. The strong-deflection formalism provides analytic expressions for the bending angle, consisting of a logarithmically divergent term together with a finite constant correction. Substituting these expressions into the lens equation allows one to obtain approximate analytic formulas for the angular positions, separations, and magnifications of the relativistic images generated in the strong-gravity region~\cite{Virbhadra:1999nm,Bozza:2002zj,Bozza:2003cp,Tsukamoto:2016jzh}.
We consider the standard gravitational lensing configuration in which a light ray emitted by a source $S$ is deflected by a compact lens $L$ before reaching an observer $O$. The source is located at an angular position $\beta$ with respect to the optical axis, while the resulting image is observed at an angular position $\theta$. The corresponding photon trajectory is characterized by an impact parameter $u$ and undergoes a total deflection angle $\alpha$. In the strong-field regime, photons may execute one or more complete loops around the compact object before escaping, giving rise to an infinite sequence of relativistic images on either side of the lens~\cite{Darwin:1959zz,Virbhadra:1999nm,Bozza:2002zj,Perlick:2021aok}.

A small-angle thin-lens approximation can be written as~\cite{Virbhadra:1999nm,Bozza:2002zj,Bozza:2008ev,Bozza:2003cp,Perlick:2021aok}
\begin{equation}\label{lensequation}
\beta=\theta-\frac{D_{LS}}{D_{OS}}\Delta\alpha_n,
\end{equation}
where $D_{LS}$ is the distance between the lens and the source, $D_{OS}$ is the observer–source distance with $D_{OS}=D_{OL}+D_{LS}$, and $D_{OL}$ is the observer–lens distance. The quantity $\Delta\alpha_n=\alpha(\theta)-2\pi n$ represents the deviation of the deflection angle after subtracting the winding contribution $2\pi n$, where $n$ counts the number of loops made by the photon around the lens.
To proceed, we define the image position $\theta_n^0$ through the condition
\begin{equation}\label{theta0n}
\alpha(\theta_n^0)=2\pi n.
\end{equation}
We then expand the deflection angle around $\theta=\theta_n^0$ as~\cite{Bozza:2002zj,Bozza:2003cp}
\begin{equation}\label{alphaexpand}
\alpha(\theta)=\alpha(\theta_n^0)+\left.\frac{d\alpha}{d\theta}\right|_{\theta=\theta_n^0}(\theta-\theta_n^0)+\mathcal{O}\!\left((\theta-\theta_n^0)^2\right).
\end{equation}

\subsection{Images slightly outside the primary photon sphere}

We follow the standard strong-deflection formalism developed in Ref.~\cite{Bozza:2002zj} to obtain the properties of relativistic images formed on the outer side of the photon sphere. Using Eqs.~(\ref{theta0n}) and (\ref{def4}), the angular position $\theta_n^0$ can be written as
\begin{equation}\label{theta0ninf}
\theta_n^0=\frac{u_m(1+e_n)}{D_{OL}}, \qquad 
e_n=\exp\left(\frac{\bar{b}-2n\pi}{\bar{a}}\right),
\end{equation}
where $u_m$ is the critical impact parameter associated with the photon sphere.
Substituting Eq.~(\ref{def4}) into Eq.~(\ref{alphaexpand}) and using the lens equation~(\ref{lensequation}), we obtain the image position as
\begin{equation}
\theta_n(\beta)=\theta_n^0-\frac{u_m e_n D_{OS}}{\bar{a}D_{LS}D_{OL}}(\theta_n^0-\beta).
\end{equation}

In the strong-field regime, the angular positions of the images decrease with increasing winding number $n$. The first relativistic image corresponds to $n=1$, while the remaining images accumulate toward a limiting angle $\theta_\infty$.

Only the outermost image is typically resolvable in practice, since the others lie very close to $\theta_\infty$ and are strongly demagnified. We therefore introduce two standard observables to characterize this structure. The first is the angular separation between the outermost image and the rest,
\begin{eqnarray}\label{outerset}
s_1 &\equiv& \theta_1-\theta_\infty \simeq \theta_\infty e_1, \nonumber\\
\mu_n &\equiv& \frac{\theta_n}{\beta}\frac{d\theta_n}{d\beta}
\simeq \frac{\theta_\infty^2 D_{OS}(1+e_n)e_n}{\bar{a}\beta D_{LS}},
\end{eqnarray}
where the approximation follows from the strong-deflection expansion.

The first relativistic image ($n=1$) is the brightest among the infinite sequence. The total magnification of all images and that excluding the outermost one are given by
\begin{eqnarray}
\sum_{n=1}^\infty \mu_n 
&\simeq& 
\frac{\theta_\infty^2 D_{OS}(1+e^{2\pi/\bar{a}}+e^{\bar{b}/\bar{a}})\,e^{\bar{b}/\bar{a}}}
{\beta \bar{a} D_{LS}(e^{4\pi/\bar{a}}-1)}, \\
\sum_{n=2}^\infty \mu_n 
&\simeq&
\frac{\theta_\infty^2 D_{OS}(e^{2\pi/\bar{a}}+e^{4\pi/\bar{a}}+e^{\bar{b}/\bar{a}})\,e^{(\bar{b}-4\pi)/\bar{a}}}
{\beta \bar{a} D_{LS}(e^{4\pi/\bar{a}}-1)}.
\end{eqnarray}
Finally, the ratio of the magnification of the outermost image to the sum of the remaining relativistic images is
\begin{equation}
\mathcal{R}_{1}
\equiv \frac{\mu_1}{\sum_{n=2}^\infty \mu_n}
\simeq 
\frac{(e^{4\pi/\bar{a}}-1)(e^{2\pi/\bar{a}}+e^{\bar{b}/\bar{a}})}
{e^{2\pi/\bar{a}}+e^{4\pi/\bar{a}}+e^{\bar{b}/\bar{a}}}.
\end{equation}

\subsection{Images slightly inside the primary photon sphere}

For images formed slightly inside the primary photon sphere, we use the deflection angle $\alpha_D(u)$ given in Eq.~(\ref{defInside}). Using Eq.~(\ref{theta0n}), the leading angular position can be written as
\begin{equation}\label{theta02}
\theta_n^0=\frac{u_m(1-e_n)}{D_{OL}}, \qquad 
e_n=\exp\left(\frac{\bar{d}-2n\pi}{\bar{c}}\right),
\end{equation}
where $u_m$ is the critical impact parameter.
Substituting Eq.~(\ref{defInside}) into Eq.~(\ref{alphaexpand}) and using the lens equation~(\ref{lensequation}), the image position becomes
\begin{equation}
\theta_n(\beta)=\theta_n^0+\frac{u_m e_n D_{OS}}{\bar{c}D_{LS}D_{OL}}(\theta_n^0-\beta).
\end{equation}

The angular separation between the limiting image position and the first image in this sequence is
\begin{equation}
s_{-1} \equiv \theta_1-\theta_\infty \simeq \theta_\infty e^{\frac{\bar{d}-2\pi}{\bar{c}}}.
\end{equation}

The magnification of the $n$-th image is approximately
\begin{equation}
\mu_{-n} \simeq \frac{\theta_\infty^2 D_{OS}\left(1- e^{\frac{\bar{d}-2\pi n}{\bar{c}}}\right)e^{\frac{\bar{d}-2\pi n}{\bar{c}}}}{\beta\,\bar{c}\,D_{LS}}.
\end{equation}

The innermost image ($n=1$) turns out to be the brightest in this family. The total magnification and the contribution excluding the innermost image can then be written as
\begin{eqnarray}
\sum_{n=1}^\infty \mu_n 
&\simeq& 
\frac{-\theta_\infty^2 D_{OS}\left(1+e^{2\pi/\bar{c}}-e^{\bar{d}/\bar{c}}\right)e^{\bar{d}/\bar{c}}}
{\beta\,\bar{c}\,D_{LS}\left(e^{4\pi/\bar{c}}-1\right)}, \\
\sum_{n=2}^\infty \mu_n 
&\simeq&
\frac{-\theta_\infty^2 D_{OS}\left(e^{2\pi/\bar{c}}+e^{4\pi/\bar{c}}-e^{\bar{d}/\bar{c}}\right)e^{(\bar{d}-4\pi)/\bar{c}}}
{\beta\,\bar{c}\,D_{LS}\left(e^{4\pi/\bar{c}}-1\right)}.
\end{eqnarray}
Finally, the ratio of the brightest image to the remaining ones is
\begin{equation}
\mathcal{R}_{-1}
\equiv \frac{\mu_1}{\sum_{n=2}^\infty \mu_n}
\simeq
\frac{\left(e^{4\pi/\bar{c}}-1\right)\left(e^{2\pi/\bar{c}}-e^{\bar{d}/\bar{c}}\right)}
{e^{2\pi/\bar{c}}+e^{4\pi/\bar{c}}-e^{\bar{d}/\bar{c}}}.
\end{equation}

\section{Discussion and Conclusion}\label{sec8}

The investigation of strong gravitational lensing by compact objects has emerged as one of the most powerful probes of the nature of gravity in the strong-field regime, particularly following the horizon-scale imaging of supermassive black holes by the EHT and the detection of GWs from binary compact object mergers. Strong lensing in black hole spacetimes has been discussed quite extensively in GR and modified theories of gravity. Still, it is not completely settled, as horizonless ultracompact configurations can also arise in several regular or modified gravities, and they may leave different observational imprints. Among these, black-bounce geometries are particularly useful since they interpolate between black-hole and wormhole-like solutions through a deformation parameter. Depending on the parameter range, one can have more than one photon sphere, with both stable and unstable null circular orbits, which already complicates the usual lensing picture compared to Kerr or Schwarzschild cases. Motivated by this, in this paper we examine strong-deflection gravitational lensing in rotating black-bounce spacetimes and seek to isolate the changes that arise when the compact object is a horizonless ultracompact object rather than a black hole.

Our analysis focuses mainly on two new black-bounce models, referred to as \textit{Model-II} and \textit{Model-III}, which extend the original Simpson--Visser geometry through different choices of the parameters $(n,k)$ in the mass function (see Table~\ref{table1} for a summary). \textit{Model-I}, which preserves the overall geometric and causal structure of the Simpson--Visser spacetime across its parameter space, has been studied previously by us \cite{Islam:2021ful}. Both \textit{Model-II} and \textit{Model-III} are regular spacetimes characterized by the bounce parameter $\ell$, which removes the central curvature singularity and controls the transition between black-hole and wormhole geometries. In the limit $\ell \rightarrow 0$, both models reduce to the Kerr black-hole solution. Since the mass functions differ for these two models, the near-throat region and the photon-sphere structure also change, making a direct comparison of strong gravitational lensing important if one wishes to distinguish them from Kerr black holes.

We began by studying the null geodesics of these spacetimes and showed that for $\ell < \ell_{E}$ the geometry corresponds to a regular black hole with a single unstable photon sphere. In contrast, for $\ell_{E} < \ell < \ell_{P}$ the spacetime is horizonless and contains two unstable photon spheres and one stable antiphoton sphere. This bifurcation in the photon-orbit structure represents a fundamental difference from Kerr black holes and gives rise to qualitatively different lensing phenomenology. For \textit{Model-II} with spin $a = 0.2M$, a single unstable circular photon orbit exists for $0 \leq \ell \leq \ell_E = 0.7523M$. Two circular photon orbits arise in the horizonless interval $\ell_E \leq \ell \leq \ell_P = 0.81M$, with the inner orbit being stable and the outer orbit unstable. Similarly, for \textit{Model-III} with spin $a = 0.2M$, a single unstable circular photon orbit exists for $0 \leq \ell \leq \ell_E = 1.043M$, whereas within the no-horizon regime $\ell_E \leq \ell \leq \ell_P = 1.198M$, an inner stable orbit and an outer unstable orbit coexist. These features are depicted in Fig.~3, which illustrates the radial evolution of photon spheres, antiphoton spheres, and event horizons for both models, revealing that \textit{Model-III} supports photon spheres over a broader parameter range than Model-II.

We find that the strong-deflection angle exhibits a logarithmic divergence not only when photon trajectories approach the photon sphere from outside, but also when they approach it from the inside. In the latter case, photons can cross the photon sphere, pass through the antiphoton sphere, and are eventually reflected by the effective potential before escaping to infinity. This leads to two distinct families of relativistic images: those formed just outside the primary photon sphere, characterized by the lensing coefficients $(\bar{a},\bar{b})$ listed in Table~\ref{table4A} for black-hole configurations and Table~\ref{table4B} for horizonless configurations, and those formed from inside the primary photon sphere, characterized by $(\bar{c},\bar{d})$ given in Tables~\ref{table4B3} and \ref{table4B4}. In particular, \textit{Model-III} yields larger values of $\bar{c}$ and $\bar{d}$ over the same parameter range, indicating stronger light deflection and more pronounced deviations from Kerr black-hole spacetimes, especially in the horizonless regime where the antiphoton sphere plays a more significant dynamical role.

The observational implications of our results are significant. For images formed inside the primary photon sphere, we find that the angular separations $s_{-1}$ and magnifications $\mu_{-n}$ can exceed those of the images formed outside by several orders of magnitude, as evidenced by the numerical results presented in Tables~\ref{Table8A} and \ref{Table8B}. Specifically, for \textit{Model-II} with $a = 0.1M$ and $\ell = 0.8M$, the magnification of the innermost image is $\mu_{-1} \sim 10^{-15}$, whereas the corresponding outer image has $\mu_{1} \sim 10^{-17}$. At the same time, the angular separation $s_{-1}$ can be as large as several $\mu{\rm as}$, potentially within the reach of future ng-EHT observations \cite{Ayzenberg:2023hfw}. The corresponding magnifications and separations for \textit{Model-III} show similar trends but with quantitative differences that may allow the two models to be distinguished observationally. The ratio $\mathcal{R}_{\pm1}$ of the brightest image to the remaining images, listed in Tables~VIII and IX, also exhibits model-dependent variations and therefore provides an additional observational discriminant. In particular, the magnifications of the innermost images in \textit{Model-III} are larger than those in \textit{Model-II} for the same parameter values, suggesting that \textit{Model-III} configurations may be easier to detect in future high-precision ng-EHT observations.

Applying our formalism to Sgr~A$^{*}$ using the EHT observational bounds, we find that the observed shadow size $\theta_{\rm sh}\in[41.7,55.7],\mu{\rm as}$ is consistent with both black-hole and horizonless configurations of our models for low spin parameters. For $a=0.1M$, \textit{Model-II} describes regular black holes for $0<\ell/M\leq0.7523$. When $\ell/M$ exceeds this value, the spacetime becomes horizonless up to $\ell/M<0.85$. Similarly, in Model-III, regular black holes exist for $0<\ell/M\leq1.055$, while horizonless solutions occur within the interval $1.055<\ell/M<1.25$. Since current EHT observations of Sgr~A$^{*}$ do not strongly support rapidly rotating black holes, the low-spin case considered here is of particular interest. Figure~1 summarizes the allowed parameter space for both models. The light-blue region corresponds to black-hole solutions, whereas the gray-shaded region represents horizonless spacetimes that still possess closed photon rings. An interesting feature of \textit{Model-III} is that it permits a substantially larger horizonless region than Model-II, thereby providing more room to satisfy current observational constraints while still exhibiting distinctive strong-lensing signatures. It should be  noted that the consistency between the black-bounce models and the EHT  observations of Sgr A* \cite{EventHorizonTelescope:2022wkp} is based primarily on the shadow-size  constraints, which were derived assuming predominantly low-spin  configurations. For higher spin values, the shadow morphology may deviate 
more significantly from the Kerr prediction, and the current EHT 
observations may provide tighter constraints. A comprehensive comparison of 
the full image morphology, including the brightness distribution of the 
photon ring, would be required to fully assess the viability of high-spin 
horizonless configurations.

We further find that the presence of closed photon-ring structures in horizonless black-bounce spacetimes differs fundamentally from Kerr naked singularities, which typically produce only open arc-like structures. Both \textit{Model-II} and \textit{Model-III} exhibit closed photon rings for appropriate parameter values (cf. Fig.~\ref{photonring}). The existence or absence of closed photon rings may therefore provide a robust observational discriminant between regular horizonless compact objects and over-spinning Kerr geometries. A comparison between the two models further reveals that \textit{Model-III} supports closed photon rings over a larger region of parameter space, reflecting differences in its mass function and near-throat geometry.

The strong gravitational lensing formalism developed here provides a framework for analyzing relativistic images in spacetimes containing multiple photon spheres, thereby generalizing the seminal work of Bozza to horizonless geometries. We demonstrate that the bounce parameter $\ell$ and the deformation parameters $(n,k)$ leave characteristic imprints on the strong-deflection coefficients and lensing observables, which can be used to test the paradigm of regular black holes and wormholes against observational data. In the non-rotating limit, we consistently recover the lensing results previously obtained by Tsukamoto for the static \textit{Model-II} spacetime, thereby validating our analytical framework and providing a nontrivial consistency check of the rotating generalization. The comparison between \textit{Model-II} and \textit{Model-III} demonstrates how sensitive strong-lensing observables are to the specific form of the mass function and highlights the importance of accurate theoretical modeling for the interpretation of astrophysical observations.

Finally, we note that the present analysis is restricted to equatorial photon motion. Nevertheless, the results for Models II and III demonstrate that strong gravitational lensing can effectively distinguish between different classes of regular compact objects. The formalism developed here can also be extended to other black-bounce and horizonless spacetimes, making it a valuable tool for studying strong-field gravity and its observational signatures.

Several directions deserve further investigation. The analysis can be extended to non-equatorial photon trajectories, potentially leading to richer image structures and higher-order photon rings. Incorporating realistic astrophysical environments, such as accretion disks, jets, and plasma effects, will also be essential for meaningful comparisons with observations, as these environments can significantly modify both shadow and lensing features. Furthermore, including realistic accretion flows and plasma propagation effects will be important for confronting theoretical predictions with observations from the EHT and future space-based missions such as BHEX. It would also be valuable to combine strong-lensing observables with gravitational-wave signals, particularly from EMRI systems, to develop a multi-messenger framework for probing compact objects. More generally, the formalism presented here can be applied to a broad class of black-bounce and horizonless spacetimes. With future improvements in image reconstruction techniques and high-resolution observations from next-generation facilities such as ngEHT and BHEX, it may become possible to distinguish black holes from horizonless alternatives and test gravity in the strong-field regime with unprecedented precision.

\section*{Acknowledgements}
S.U.I. and S.G.G. gratefully acknowledge the warm hospitality extended by the ITPC during the course of this work. This work is supported by National Natural Science Foundation of China under Grants No. 12275238, No. 12542053, and No. W2433018, and the National Key Research and Development Program of China under Grant No. 2020YFC2201503. 
\bibliography{BBLens}

@article{LIGOScientific:2016lio,
    author = "Abbott, B. P. and others",
    collaboration = "LIGO Scientific, Virgo",
    title = "{Tests of general relativity with GW150914}",
    eprint = "1602.03841",
    archivePrefix = "arXiv",
    primaryClass = "gr-qc",
    reportNumber = "LIGO-P1500213",
    doi = "10.1103/PhysRevLett.116.221101",
    journal = "Phys. Rev. Lett.",
    volume = "116",
    number = "22",
    pages = "221101",
    year = "2016",
    note = "[Erratum: Phys.Rev.Lett. 121, 129902 (2018)]"
}

@article{EventHorizonTelescope:2019dse,
    author = "Akiyama, Kazunori and others",
    collaboration = "Event Horizon Telescope",
    title = "{First M87 Event Horizon Telescope Results. I. The Shadow of the Supermassive Black Hole}",
    eprint = "1906.11238",
    archivePrefix = "arXiv",
    primaryClass = "astro-ph.GA",
    doi = "10.3847/2041-8213/ab0ec7",
    journal = "Astrophys. J. Lett.",
    volume = "875",
    pages = "L1",
    year = "2019"
}

@article{Ashtekar:2004eh,
    author = "Ashtekar, Abhay and Lewandowski, Jerzy",
    title = "{Background independent quantum gravity: A Status report}",
    eprint = "gr-qc/0404018",
    archivePrefix = "arXiv",
    doi = "10.1088/0264-9381/21/15/R01",
    journal = "Class. Quant. Grav.",
    volume = "21",
    pages = "R53",
    year = "2004"
}

@article{Kiefer:2013jqa,
    author = "Kiefer, Claus",
    title = "{Conceptual Problems in Quantum Gravity and Quantum Cosmology}",
    eprint = "1401.3578",
    archivePrefix = "arXiv",
    primaryClass = "gr-qc",
    doi = "10.1155/2013/509316",
    journal = "ISRN Math. Phys.",
    volume = "2013",
    pages = "509316",
    year = "2013"
}

@article{SupernovaSearchTeam:1998fmf,
    author = "Riess, Adam G. and others",
    collaboration = "Supernova Search Team",
    title = "{Observational evidence from supernovae for an accelerating universe and a cosmological constant}",
    eprint = "astro-ph/9805201",
    archivePrefix = "arXiv",
    doi = "10.1086/300499",
    journal = "Astron. J.",
    volume = "116",
    pages = "1009--1038",
    year = "1998"
}

@article{Sotiriou:2008rp,
    author = "Sotiriou, Thomas P. and Faraoni, Valerio",
    title = "{f(R) Theories Of Gravity}",
    eprint = "0805.1726",
    archivePrefix = "arXiv",
    primaryClass = "gr-qc",
    doi = "10.1103/RevModPhys.82.451",
    journal = "Rev. Mod. Phys.",
    volume = "82",
    pages = "451--497",
    year = "2010"
}

@article{Clifton:2011jh,
    author = "Clifton, Timothy and Ferreira, Pedro G. and Padilla, Antonio and Skordis, Constantinos",
    title = "{Modified Gravity and Cosmology}",
    eprint = "1106.2476",
    archivePrefix = "arXiv",
    primaryClass = "astro-ph.CO",
    doi = "10.1016/j.physrep.2012.01.001",
    journal = "Phys. Rept.",
    volume = "513",
    pages = "1--189",
    year = "2012"
}

@article{Joyce:2014kja,
    author = "Joyce, Austin and Jain, Bhuvnesh and Khoury, Justin and Trodden, Mark",
    title = "{Beyond the Cosmological Standard Model}",
    eprint = "1407.0059",
    archivePrefix = "arXiv",
    primaryClass = "astro-ph.CO",
    doi = "10.1016/j.physrep.2014.12.002",
    journal = "Phys. Rept.",
    volume = "568",
    pages = "1--98",
    year = "2015"
}

@article{Nojiri:2017ncd,
    author = "Nojiri, S. and Odintsov, S. D. and Oikonomou, V. K.",
    title = "{Modified Gravity Theories on a Nutshell: Inflation, Bounce and Late-time Evolution}",
    eprint = "1705.11098",
    archivePrefix = "arXiv",
    primaryClass = "gr-qc",
    reportNumber = "PHYS.REPT.-692-(2017)-1-104, Phys.Rept. 692 (2017) 1-104",
    doi = "10.1016/j.physrep.2017.06.001",
    journal = "Phys. Rept.",
    volume = "692",
    pages = "1--104",
    year = "2017"
}

@article{Biswas:2005qr,
    author = "Biswas, Tirthabir and Mazumdar, Anupam and Siegel, Warren",
    title = "{Bouncing universes in string-inspired gravity}",
    eprint = "hep-th/0508194",
    doi = "10.1088/1475-7516/2006/03/009",
    journal = "JCAP",
    volume = "3",
    pages = "9",
    year = "2006"
}

@article{Cardoso:2016rao,
    author = "Cardoso, Vitor and Franzin, Edgardo and Pani, Paolo",
    title = "{Is the gravitational-wave ringdown a probe of the event horizon?}",
    eprint = "1602.07309",
    archivePrefix = "arXiv",
    primaryClass = "gr-qc",
    doi = "10.1103/PhysRevLett.116.171101",
    journal = "Phys. Rev. Lett.",
    volume = "116",
    number = "17",
    pages = "171101",
    year = "2016",
    note = "[Erratum: Phys.Rev.Lett. 117, 089902 (2016)]"
}

@article{Cunha:2018gql,
    author = "Cunha, Pedro V. P. and Herdeiro, Carlos A. R. and Rodriguez, Maria J.",
    title = "{Does the black hole shadow probe the event horizon geometry?}",
    eprint = "1802.02675",
    archivePrefix = "arXiv",
    primaryClass = "gr-qc",
    doi = "10.1103/PhysRevD.97.084020",
    journal = "Phys. Rev. D",
    volume = "97",
    number = "8",
    pages = "084020",
    year = "2018"
}

@article{Carballo-Rubio:2018jzw,
    author = "Carballo-Rubio, Ra{\'u}l and Di Filippo, Francesco and Liberati, Stefano and Visser, Matt",
    title = "{Phenomenological aspects of black holes beyond general relativity}",
    eprint = "1809.08238",
    archivePrefix = "arXiv",
    primaryClass = "gr-qc",
    doi = "10.1103/PhysRevD.98.124009",
    journal = "Phys. Rev. D",
    volume = "98",
    number = "12",
    pages = "124009",
    year = "2018"
}

@article{Konoplya:2018yrp,
    author = "Konoplya, R. A. and Stuchl{\'\i}k, Z. and Zhidenko, A.",
    title = "{Echoes of compact objects: new physics near the surface and matter at a distance}",
    eprint = "1810.01295",
    archivePrefix = "arXiv",
    primaryClass = "gr-qc",
    doi = "10.1103/PhysRevD.99.024007",
    journal = "Phys. Rev. D",
    volume = "99",
    number = "2",
    pages = "024007",
    year = "2019"
}

@article{Hayward:2005gi,
    author = "Hayward, Sean A.",
    title = "{Formation and evaporation of regular black holes}",
    eprint = "gr-qc/0506126",
    archivePrefix = "arXiv",
    doi = "10.1103/PhysRevLett.96.031103",
    journal = "Phys. Rev. Lett.",
    volume = "96",
    pages = "031103",
    year = "2006"
}

@article{AyonBeato:2000zs,
    author = "Ayon-Beato, Eloy and Garcia, Alberto",
    title = "{The Bardeen model as a nonlinear magnetic monopole}",
    eprint = "gr-qc/0009077",
    archivePrefix = "arXiv",
    doi = "10.1016/S0370-2693(00)01125-4",
    journal = "Phys. Lett. B",
    volume = "493",
    pages = "149--152",
    year = "2000"
}

@article{AyonBeato:1998ub,
    author = "Ayon-Beato, Eloy and Garcia, Alberto",
    title = "{Regular black hole in general relativity coupled to nonlinear electrodynamics}",
    eprint = "gr-qc/9911046",
    archivePrefix = "arXiv",
    doi = "10.1103/PhysRevLett.80.5056",
    journal = "Phys. Rev. Lett.",
    volume = "80",
    pages = "5056--5059",
    year = "1998"
}

@article{KumarWalia:2022aop,
    author = "Kumar Walia, Rahul and Ghosh, Sushant G. and Maharaj, Sunil D.",
    title = "{Testing Rotating Regular Metrics with EHT Results of Sgr A*}",
    eprint = "2207.00078",
    archivePrefix = "arXiv",
    primaryClass = "gr-qc",
    doi = "10.3847/1538-4357/ac9623",
    journal = "Astrophys. J.",
    volume = "939",
    number = "2",
    pages = "77",
    year = "2022"
}

@article{Kumar:2020yem,
    author = "Kumar, Rahul and Kumar, Amit and Ghosh, Sushant G.",
    title = "{Testing Rotating Regular Metrics as Candidates for Astrophysical Black Holes}",
    eprint = "2006.09869",
    archivePrefix = "arXiv",
    primaryClass = "gr-qc",
    doi = "10.3847/1538-4357/ab8c4a",
    journal = "Astrophys. J.",
    volume = "896",
    number = "1",
    pages = "89",
    year = "2020"
}

@article{Yang:2022xxh,
    author = {Yang, Yi and Liu, Dong and {\"O}vg{\"u}n, Ali and Long, Zheng-Wen and Xu, Zhaoyi},
    title = "{Quasinormal modes of Kerr-like black bounce spacetime}",
    eprint = "2205.07530",
    archivePrefix = "arXiv",
    journal = "",
    volume = "",
    number = "",
    primaryClass = "gr-qc",
    month = "5",
    year = "2022"
}

@article{Bambhaniya:2021ugr,
    author = "Bambhaniya, Parth and K, Saurabh and Jusufi, Kimet and Joshi, Pankaj S.",
    title = "{Thin accretion disk in the Simpson-Visser black-bounce and wormhole spacetimes}",
    eprint = "2109.15054",
    archivePrefix = "arXiv",
    primaryClass = "gr-qc",
    doi = "10.1103/PhysRevD.105.023021",
    journal = "Phys. Rev. D",
    volume = "105",
    number = "2",
    pages = "023021",
    year = "2022"
}

@article{Kumar:2022fqo,
    author = "Kumar, Jitendra and Islam, Shafqat Ul and Ghosh, Sushant G.",
    title = "{Testing Strong Gravitational Lensing Effects of Supermassive Compact Objects with Regular Spacetimes}",
    eprint = "2209.04240",
    archivePrefix = "arXiv",
    primaryClass = "gr-qc",
    doi = "10.3847/1538-4357/ac912c",
    journal = "Astrophys. J.",
    volume = "938",
    number = "2",
    pages = "104",
    year = "2022"
}

@article{Bardeen:1968,
    author = "Bardeen, J. M.",
    title = "{Circular geodesic of Bardeen and Ayon\textendash{}Beato\textendash{}Garcia regular black-hole and no-horizon spacetimes}",
    doi = "10.1142/S0218271815500200",
    journal = " in Conference Proceedings of GR5",
    pages = "174",
    year = "1968"
}

@article{Akiyama:2019cqa,
  author  = {Akiyama, K. and others},
  collaboration = {Event Horizon Telescope},
  title   = {First M87 Event Horizon Telescope Results. I. The Shadow of the Supermassive Black Hole},
  journal = {Astrophys. J.},
  volume  = {875},
  pages   = {L1},
  year    = {2019}
}

@article{Perlick:2021aok,
    author = "Perlick, Volker and Tsupko, Oleg Yu.",
    title = "{Calculating black hole shadows: Review of analytical studies}",
    eprint = "2105.07101",
    archivePrefix = "arXiv",
    primaryClass = "gr-qc",
    doi = "10.1016/j.physrep.2021.10.004",
    journal = "Phys. Rept.",
    volume = "947",
    pages = "1--39",
    year = "2022"
}

@article{Simpson:2019mud,
    author = "Simpson, Alex and Visser, Matt",
    title = "{Regular black holes with asymptotically Minkowski cores}",
    eprint = "1911.01020",
    archivePrefix = "arXiv",
    primaryClass = "gr-qc",
    doi = "10.3390/universe6010008",
    journal = "Universe",
    volume = "6",
    number = "1",
    pages = "8",
    year = "2019"
}

@article{Carter:1968rr,
    author = "Carter, Brandon",
    title = "{Global structure of the Kerr family of gravitational fields}",
    doi = "10.1103/PhysRev.174.1559",
    journal = "Phys. Rev.",
    volume = "174",
    pages = "1559--1571",
    year = "1968"
}

@article{Bardeen:1972fi,
    author = "Bardeen, James M. and Press, William H. and Teukolsky, Saul A",
    title = "{Rotating black holes: Locally nonrotating frames, energy extraction, and scalar synchrotron radiation}",
    reportNumber = "OAP-288",
    doi = "10.1086/151796",
    journal = "Astrophys. J.",
    volume = "178",
    pages = "347",
    year = "1972"
}

@article{Frolov:2016pav,
    author = "Frolov, Valeri P.",
    title = "{Notes on nonsingular models of black holes}",
    eprint = "1609.01758",
    archivePrefix = "arXiv",
    primaryClass = "gr-qc",
    doi = "10.1103/PhysRevD.94.104056",
    journal = "Phys. Rev. D",
    volume = "94",
    number = "10",
    pages = "104056",
    year = "2016"
}

@article{Perlick:2004tq,
    author = "Perlick, V.",
    title = "{Gravitational lensing from a spacetime perspective}",
    doi = "10.12942/lrr-2004-9",
    journal = "Living Rev. Rel.",
    volume = "7",
    pages = "9",
    year = "2004"
}

@article{Modesto:2017sdr,
    author = "Modesto, Leonardo and Rachwa{\l}, Les{\l}aw",
    title = "{Nonlocal quantum gravity: A review}",
    doi = "10.1142/S0218271817300208",
    journal = "Int. J. Mod. Phys. D",
    volume = "26",
    number = "11",
    pages = "1730020",
    year = "2017"
}

@article{Bronnikov:2021uta,
    author = "Bronnikov, Kirill A. and Walia, Rahul Kumar",
    title = "{Field sources for Simpson-Visser spacetimes}",
    eprint = "2112.13198",
    archivePrefix = "arXiv",
    primaryClass = "gr-qc",
    doi = "10.1103/PhysRevD.105.044039",
    journal = "Phys. Rev. D",
    volume = "105",
    number = "4",
    pages = "044039",
    year = "2022"
}

@article{Grenzebach:2014fha,
    author = {Grenzebach, Arne and Perlick, Volker and L{\"a}mmerzahl, Claus},
    title = "{Photon Regions and Shadows of Kerr-Newman-NUT Black Holes with a Cosmological Constant}",
    eprint = "1403.5234",
    archivePrefix = "arXiv",
    primaryClass = "gr-qc",
    doi = "10.1103/PhysRevD.89.124004",
    journal = "Phys. Rev. D",
    volume = "89",
    number = "12",
    pages = "124004",
    year = "2014"
}

@article{Johannsen:2010ru,
    author = "Johannsen, Tim and Psaltis, Dimitrios",
    title = "{Testing the No-Hair Theorem with Observations in the Electromagnetic Spectrum: II. Black-Hole Images}",
    eprint = "1005.1931",
    archivePrefix = "arXiv",
    primaryClass = "astro-ph.HE",
    doi = "10.1088/0004-637X/718/1/446",
    journal = "Astrophys. J.",
    volume = "718",
    pages = "446--454",
    year = "2010"
}

@article{Synge:1966okc,
    author = "Synge, J. L.",
    title = "{The Escape of Photons from Gravitationally Intense Stars}",
    doi = "10.1093/mnras/131.3.463",
    journal = "Mon. Not. Roy. Astron. Soc.",
    volume = "131",
    number = "3",
    pages = "463--466",
    year = "1966"
}

@article{Bambi:2010hf,
    author = "Bambi, Cosimo and Yoshida, Naoki",
    title = "{Shape and position of the shadow in the $\delta = 2$ Tomimatsu-Sato space-time}",
    eprint = "1004.3149",
    archivePrefix = "arXiv",
    primaryClass = "gr-qc",
    reportNumber = "IPMU10-0061",
    doi = "10.1088/0264-9381/27/20/205006",
    journal = "Class. Quant. Grav.",
    volume = "27",
    pages = "205006",
    year = "2010"
}

@article{Stuchlik:2019uvf,
    author = "Stuchl{\'\i}k, Zdenek and Schee, Jan",
    title = "{Shadow of the regular Bardeen black holes and comparison of the motion of photons and neutrinos}",
    doi = "10.1140/epjc/s10052-019-6543-8",
    journal = "Eur. Phys. J. C",
    volume = "79",
    number = "1",
    pages = "44",
    year = "2019"
}

@article{Kumar:2020ltt,
    author = "Kumar, Rahul and Ghosh, Sushant G.",
    title = "{Photon ring structure of rotating regular black holes and no-horizon spacetimes}",
    eprint = "2004.07501",
    archivePrefix = "arXiv",
    primaryClass = "gr-qc",
    doi = "10.1088/1361-6382/abdd48",
    journal = "Class. Quant. Grav.",
    volume = "38",
    number = "8",
    pages = "8",
    year = "2021"
}

@article{Cunha:2018acu,
    author = "Cunha, Pedro V. P. and Herdeiro, Carlos A. R.",
    title = "{Shadows and strong gravitational lensing: a brief review}",
    eprint = "1801.00860",
    archivePrefix = "arXiv",
    primaryClass = "gr-qc",
    doi = "10.1007/s10714-018-2361-9",
    journal = "Gen. Rel. Grav.",
    volume = "50",
    number = "4",
    pages = "42",
    year = "2018"
}

@article{Grenzebach:2015oea,
    author = {Grenzebach, Arne and Perlick, Volker and L{\"a}mmerzahl, Claus},
    editor = "Herdeiro, Carlos A. R. and Cardoso, Vitor and Lemos, Jose P. S. and Mena, Filipe C.",
    title = "{Photon Regions and Shadows of Accelerated Black Holes}",
    eprint = "1503.03036",
    archivePrefix = "arXiv",
    primaryClass = "gr-qc",
    doi = "10.1142/S0218271815420249",
    journal = "Int. J. Mod. Phys. D",
    volume = "24",
    number = "09",
    pages = "1542024",
    year = "2015"
}

@article{Newman:1965tw,
    author = "Newman, E. T. and Janis, A. I.",
    title = "{Note on the Kerr spinning particle metric}",
    doi = "10.1063/1.1704350",
    journal = "J. Math. Phys.",
    volume = "6",
    pages = "915--917",
    year = "1965"
}

@article{Falcke:1999pj,
    author = "Falcke, Heino and Melia, Fulvio and Agol, Eric",
    title = "{Viewing the shadow of the black hole at the galactic center}",
    eprint = "astro-ph/9912263",
    archivePrefix = "arXiv",
    reportNumber = "HFA-EPRINT-NO-33",
    doi = "10.1086/312423",
    journal = "Astrophys. J. Lett.",
    volume = "528",
    pages = "L13",
    year = "2000"
}

@article{Hayward:2006ej,
    author = "Hayward, Sean A.",
    title = "{Formation and evaporation of regular black holes}",
    eprint = "gr-qc/0506126",
    archivePrefix = "arXiv",
    doi = "10.1103/PhysRevLett.96.031103",
    journal = "Phys. Rev. Lett.",
    volume = "96",
    pages = "031103",
    year = "2006"
}

@article{Ghosh:2022mka,
    author = "Ghosh, Saptaswa and Bhattacharyya, Arpan",
    title = "{Analytical study of gravitational lensing in Kerr-Newman black-bounce spacetime}",
    eprint = "2206.09954",
    archivePrefix = "arXiv",
    primaryClass = "gr-qc",
    doi = "10.1088/1475-7516/2022/11/006",
    journal = "JCAP",
    volume = "11",
    pages = "006",
    year = "2022"
}

@article{Modesto:2004xx,
    author = "Modesto, Leonardo",
    title = "{Disappearance of black hole singularity in quantum gravity}",
    eprint = "gr-qc/0407097",
    archivePrefix = "arXiv",
    doi = "10.1103/PhysRevD.70.124009",
    journal = "Phys. Rev. D",
    volume = "70",
    pages = "124009",
    year = "2004"
}

@book{Chandrasekhar:1985kt,
    author = "Chandrasekhar, Subrahmanyan",
    title = "{The mathematical theory of black holes}",
    isbn = "978-0-19-850370-5",
    year = "1985"
}

@article{Perlick:2010zh,
    author = "Perlick, Volker",
    title = "Gravitational Lensing from a Spacetime Perspective",
    eprint = "1010.3416",
    journal = "arXiv",
    archivePrefix = "arXiv",
    primaryClass = "gr-qc",
    month = "10",
    year = "2010"
}

@article{Peng:2018nkj,
    author = "Peng, Yan",
    title = "{Upper bound on the radii of regular ultra-compact star photonspheres}",
    eprint = "1812.04257",
    archivePrefix = "arXiv",
    primaryClass = "gr-qc",
    doi = "10.1016/j.physletb.2019.01.049",
    journal = "Phys. Lett. B",
    volume = "790",
    pages = "396--399",
    year = "2019"
}

@article{Hod:2020pim,
    author = "Hod, Shahar",
    title = "{Lower bound on the radii of black-hole photonspheres}",
    eprint = "2012.03962",
    archivePrefix = "arXiv",
    primaryClass = "gr-qc",
    doi = "10.1103/PhysRevD.101.084033",
    journal = "Phys. Rev. D",
    volume = "101",
    number = "8",
    pages = "084033",
    year = "2020"
}

@article{Hod:2023jmx,
    author = "Hod, Shahar",
    title = "{Lower bound on the radii of light rings in traceless black-hole spacetimes}",
    eprint = "2311.17462",
    archivePrefix = "arXiv",
    primaryClass = "gr-qc",
    doi = "10.1007/JHEP12(2023)178",
    journal = "JHEP",
    volume = "12",
    pages = "178",
    year = "2023"
}

@article{Hod:2017zpi,
    author = "Hod, Shahar",
    title = "{On the number of light rings in curved spacetimes of ultra-compact objects}",
    eprint = "1710.00836",
    archivePrefix = "arXiv",
    primaryClass = "gr-qc",
    doi = "10.1016/j.physletb.2017.11.021",
    journal = "Phys. Lett. B",
    volume = "776",
    pages = "1--4",
    year = "2018"
}

@article{Cunha:2020azh,
    author = "Cunha, Pedro V. P. and Herdeiro, Carlos A. R.",
    title = "{Stationary black holes and light rings}",
    eprint = "2003.06445",
    archivePrefix = "arXiv",
    primaryClass = "gr-qc",
    doi = "10.1103/PhysRevLett.124.181101",
    journal = "Phys. Rev. Lett.",
    volume = "124",
    number = "18",
    pages = "181101",
    year = "2020"
}

@article{Cunha:2024ajc,
    author = "Cunha, Pedro V. P. and Herdeiro, Carlos A. R. and Novo, Jo{\~a}o P. A.",
    title = "{Light rings on stationary axisymmetric spacetimes: Blind to the topology and able to coexist}",
    eprint = "2401.05495",
    archivePrefix = "arXiv",
    primaryClass = "gr-qc",
    doi = "10.1103/PhysRevD.109.064050",
    journal = "Phys. Rev. D",
    volume = "109",
    number = "6",
    pages = "064050",
    year = "2024"
}

@article{Padhye:2024yrj,
    author = "Padhye, Prasad and Paithankar, Kajol and Kolekar, Sanved",
    title = "{Photon region boundary for stationary axisymmetric black holes}",
    eprint = "2412.12665",
    archivePrefix = "arXiv",
    primaryClass = "gr-qc",
    doi = "10.1103/b84x-ch1m",
    journal = "Phys. Rev. D",
    volume = "112",
    number = "12",
    pages = "124025",
    year = "2025"
}

@article{Sanchez:1977si,
    author = "Sanchez, Norma G.",
    title = "{Absorption and Emission Spectra of a Schwarzschild Black Hole}",
    reportNumber = "Print-77-0242 (MEUDON)",
    doi = "10.1103/PhysRevD.18.1030",
    journal = "Phys. Rev. D",
    volume = "18",
    pages = "1030",
    year = "1978"
}

@article{Decanini:2010fz,
    author = "Decanini, Yves and Folacci, Antoine and Raffaelli, Bernard",
    title = "{Unstable circular null geodesics of static spherically symmetric black holes, Regge poles and quasinormal frequencies}",
    eprint = "1002.0121",
    archivePrefix = "arXiv",
    primaryClass = "gr-qc",
    doi = "10.1103/PhysRevD.81.104039",
    journal = "Phys. Rev. D",
    volume = "81",
    pages = "104039",
    year = "2010"
}

@article{Wei:2011zw,
    author = "Wei, Shao-Wen and Liu, Yu-Xiao and Guo, Heng",
    title = "{Relationship between High-Energy Absorption Cross Section and Strong Gravitational Lensing for Black Hole}",
    eprint = "1103.3822",
    archivePrefix = "arXiv",
    primaryClass = "hep-th",
    doi = "10.1103/PhysRevD.84.041501",
    journal = "Phys. Rev. D",
    volume = "84",
    pages = "041501",
    year = "2011"
}

@article{Press:1971wr,
    author = "Press, William H.",
    title = "{Long Wave Trains of Gravitational Waves from a Vibrating Black Hole}",
    reportNumber = "OAP-262",
    doi = "10.1086/180849",
    journal = "Astrophys. J. Lett.",
    volume = "170",
    pages = "L105--L108",
    year = "1971"
}

@article{Stefanov:2010xz,
    author = "Stefanov, Ivan Zh. and Yazadjiev, Stoytcho S. and Gyulchev, Galin G.",
    title = "{Connection between Black-Hole Quasinormal Modes and Lensing in the Strong Deflection Limit}",
    eprint = "1003.1609",
    archivePrefix = "arXiv",
    primaryClass = "gr-qc",
    doi = "10.1103/PhysRevLett.104.251103",
    journal = "Phys. Rev. Lett.",
    volume = "104",
    pages = "251103",
    year = "2010"
}

@article{Raffaelli:2014ola,
    author = "Raffaelli, Bernard",
    title = "{Strong gravitational lensing and black hole quasinormal modes: Towards a semiclassical unified description}",
    eprint = "1412.7333",
    archivePrefix = "arXiv",
    primaryClass = "gr-qc",
    doi = "10.1007/s10714-016-2016-7",
    journal = "Gen. Rel. Grav.",
    volume = "48",
    number = "2",
    pages = "16",
    year = "2016"
}

@article{Shaikh:2019jfr,
    author = "Shaikh, Rajibul and Banerjee, Pritam and Paul, Suvankar and Sarkar, Tapobrata",
    title = "{Strong gravitational lensing by wormholes}",
    eprint = "1905.06932",
    archivePrefix = "arXiv",
    primaryClass = "gr-qc",
    doi = "10.1088/1475-7516/2019/07/028",
    journal = "JCAP",
    volume = "07",
    pages = "028",
    year = "2019",
    note = "[Erratum: JCAP 12, E01 (2023)]"
}

@article{Gan:2021pwu,
    author = "Gan, Qingyu and Wang, Peng and Wu, Houwen and Yang, Haitang",
    title = "{Photon spheres and spherical accretion image of a hairy black hole}",
    eprint = "2104.08703",
    archivePrefix = "arXiv",
    primaryClass = "gr-qc",
    doi = "10.1103/PhysRevD.104.024003",
    journal = "Phys. Rev. D",
    volume = "104",
    number = "2",
    pages = "024003",
    year = "2021"
}

@article{Gan:2021xdl,
    author = "Gan, Qingyu and Wang, Peng and Wu, Houwen and Yang, Haitang",
    title = "{Photon ring and observational appearance of a hairy black hole}",
    eprint = "2105.11770",
    archivePrefix = "arXiv",
    primaryClass = "gr-qc",
    doi = "10.1103/PhysRevD.104.044049",
    journal = "Phys. Rev. D",
    volume = "104",
    number = "4",
    pages = "044049",
    year = "2021"
}

@article{Wang:2020emr,
    author = "Wang, Xiaobao and Li, Peng-Cheng and Zhang, Cheng-Yong and Guo, Minyong",
    title = "{Novel shadows from the asymmetric thin-shell wormhole}",
    eprint = "2007.03327",
    archivePrefix = "arXiv",
    primaryClass = "gr-qc",
    doi = "10.1016/j.physletb.2020.135930",
    journal = "Phys. Lett. B",
    volume = "811",
    pages = "135930",
    year = "2020"
}

@article{Wielgus:2020uqz,
    author = "Wielgus, Maciek and Horak, Jiri and Vincent, Frederic and Abramowicz, Marek",
    title = "{Reflection-asymmetric wormholes and their double shadows}",
    eprint = "2008.10130",
    archivePrefix = "arXiv",
    primaryClass = "gr-qc",
    doi = "10.1103/PhysRevD.102.084044",
    journal = "Phys. Rev. D",
    volume = "102",
    number = "8",
    pages = "084044",
    year = "2020"
}

@article{Tsukamoto:2021fpp,
    author = "Tsukamoto, Naoki",
    title = "{Linearization stability of reflection-asymmetric thin-shell wormholes with double shadows}",
    eprint = "2101.07060",
    archivePrefix = "arXiv",
    primaryClass = "gr-qc",
    doi = "10.1103/PhysRevD.103.064031",
    journal = "Phys. Rev. D",
    volume = "103",
    number = "6",
    pages = "064031",
    year = "2021"
}

@article{Guerrero:2021pxt,
    author = "Guerrero, Merce and Olmo, Gonzalo J. and Rubiera-Garcia, Diego",
    title = "{Double shadows of reflection-asymmetric wormholes supported by positive energy thin-shells}",
    eprint = "2102.00840",
    archivePrefix = "arXiv",
    primaryClass = "gr-qc",
    doi = "10.1088/1475-7516/2021/04/066",
    journal = "JCAP",
    volume = "04",
    pages = "066",
    year = "2021"
}

@article{Peng:2021osd,
    author = "Peng, Jun and Guo, Minyong and Feng, Xing-Hui",
    title = "{Observational signature and additional photon rings of an asymmetric thin-shell wormhole}",
    eprint = "2102.05488",
    archivePrefix = "arXiv",
    primaryClass = "gr-qc",
    doi = "10.1103/PhysRevD.104.124010",
    journal = "Phys. Rev. D",
    volume = "104",
    number = "12",
    pages = "124010",
    year = "2021"
}

@article{Igata:2025plb,
    author = "Igata, Takahisa",
    title = "{Deflection angle in the strong deflection limit for static and axisymmetric spacetimes: Local curvature, matter fields, and quasinormal modes}",
    eprint = "2504.07906",
    archivePrefix = "arXiv",
    primaryClass = "gr-qc",
    doi = "10.1103/ylrj-rm9j",
    journal = "Phys. Rev. D",
    volume = "113",
    number = "2",
    pages = "024036",
    year = "2026"
}

@article{Barcelo:2000ta,
    author = "Barcelo, C. and Visser, Matt",
    title = "{Brane surgery: Energy conditions, traversable wormholes, and voids}",
    eprint = "hep-th/0004022",
    archivePrefix = "arXiv",
    doi = "10.1016/S0550-3213(00)00379-5",
    journal = "Nucl. Phys. B",
    volume = "584",
    pages = "415--435",
    year = "2000"
}

@article{Koga:2020gqd,
    author = "Koga, Yasutaka",
    title = "{Photon surfaces as pure tension shells: Uniqueness of thin shell wormholes}",
    eprint = "2003.10859",
    archivePrefix = "arXiv",
    primaryClass = "gr-qc",
    reportNumber = "RUP-20-10",
    doi = "10.1103/PhysRevD.101.104022",
    journal = "Phys. Rev. D",
    volume = "101",
    number = "10",
    pages = "104022",
    year = "2020"
}

@article{Tsukamoto:2023kvk,
    author = "Tsukamoto, Naoki and Kokubu, Takafumi",
    title = "{Relation between circular photon orbits and the stability of wormholes with the thin shell of a barotropic fluid}",
    eprint = "2305.07256",
    archivePrefix = "arXiv",
    primaryClass = "gr-qc",
    doi = "10.1103/PhysRevD.109.024047",
    journal = "Phys. Rev. D",
    volume = "109",
    number = "2",
    pages = "024047",
    year = "2024"
}

@article{Islam:2021ful,
    author = "Islam, Shafqat Ul and Kumar, Jitendra and Ghosh, Sushant G.",
    title = "{Strong gravitational lensing by
rotating Simpson-Visser black holes}",
    eprint = "2104.00696",
    archivePrefix = "arXiv",
    primaryClass = "gr-qc",
    doi = "10.1088/1475-7516/2021/10/013",
    journal = "JCAP",
    volume = "10",
    pages = "013",
    year = "2021"
}

@article{Hod:2017xkz,
  author  = {Hod, Shahar},
  title   = {Upper bound on the radii of black-hole photonspheres},
  journal = {Phys. Lett. B},
  volume  = {727},
  pages   = {345},
  year    = {2013}
}

@article{Bronnikov:2000vy,
    author = "Bronnikov, Kirill A.",
    title = "{Regular magnetic black holes and monopoles from nonlinear electrodynamics}",
    eprint = "gr-qc/0006014",
    archivePrefix = "arXiv",
    doi = "10.1103/PhysRevD.63.044005",
    journal = "Phys. Rev. D",
    volume = "63",
    pages = "044005",
    year = "2001"
}

@article{Fan:2016hvf,
    author = "Fan, Zhong-Ying and Wang, Xiaobao",
    title = "{Construction of Regular Black Holes in General Relativity}",
    eprint = "1610.02636",
    archivePrefix = "arXiv",
    primaryClass = "gr-qc",
    doi = "10.1103/PhysRevD.94.124027",
    journal = "Phys. Rev. D",
    volume = "94",
    number = "12",
    pages = "124027",
    year = "2016"
}

@article{Toshmatov:2018cks,
    author = "Toshmatov, Bobir and Stuchl{\'\i}k, Zden{\v{e}}k and Ahmedov, Bobomurat",
    title = "{Comment on {\textquotedblleft}Construction of regular black holes in general relativity{\textquotedblright}}",
    eprint = "1807.09502",
    archivePrefix = "arXiv",
    primaryClass = "gr-qc",
    doi = "10.1103/PhysRevD.98.028501",
    journal = "Phys. Rev. D",
    volume = "98",
    number = "2",
    pages = "028501",
    year = "2018"
}

@article{Capozziello:2013vna,
    author = "Capozziello, Salvatore and Lobo, Francisco S. N. and Mimoso, Jos{\'e} P.",
    title = "{Energy conditions in modified gravity}",
    eprint = "1312.0784",
    archivePrefix = "arXiv",
    primaryClass = "gr-qc",
    doi = "10.1016/j.physletb.2014.01.066",
    journal = "Phys. Lett. B",
    volume = "730",
    pages = "280--283",
    year = "2014"
}

@article{Capozziello:2014bqa,
    author = "Capozziello, Salvatore and Lobo, Francisco S. N. and Mimoso, Jos{\'e} P.",
    title = "{Generalized energy conditions in Extended Theories of Gravity}",
    eprint = "1407.7293",
    archivePrefix = "arXiv",
    primaryClass = "gr-qc",
    doi = "10.1103/PhysRevD.91.124019",
    journal = "Phys. Rev. D",
    volume = "91",
    number = "12",
    pages = "124019",
    year = "2015"
}

@article{Rodrigues:2015ayd,
    author = "Rodrigues, Manuel E. and Junior, Ednaldo L. B. and Marques, Glauber T. and Zanchin, Vilson T.",
    title = "{Regular black holes in $f(R)$ gravity coupled to nonlinear electrodynamics}",
    eprint = "1511.00569",
    archivePrefix = "arXiv",
    primaryClass = "gr-qc",
    doi = "10.1103/PhysRevD.94.024062",
    journal = "Phys. Rev. D",
    volume = "94",
    number = "2",
    pages = "024062",
    year = "2016",
    note = "[Addendum: Phys.Rev.D 94, 049904 (2016)]"
}

@article{Rodrigues:2018bdc,
    author = "Rodrigues, Manuel E. and de Sousa Silva, Marcos V.",
    title = "{Bardeen Regular Black Hole With an Electric Source}",
    eprint = "1802.05095",
    archivePrefix = "arXiv",
    primaryClass = "gr-qc",
    doi = "10.1088/1475-7516/2018/06/025",
    journal = "JCAP",
    volume = "06",
    pages = "025",
    year = "2018"
}

@article{Bambi:2013ufa,
    author = "Bambi, Cosimo and Modesto, Leonardo",
    title = "{Rotating regular black holes}",
    eprint = "1302.6075",
    archivePrefix = "arXiv",
    primaryClass = "gr-qc",
    doi = "10.1016/j.physletb.2013.03.025",
    journal = "Phys. Lett. B",
    volume = "721",
    pages = "329--334",
    year = "2013"
}

@article{Neves:2014aba,
    author = "Neves, J. C. S. and Saa, Alberto",
    title = "{Regular rotating black holes and the weak energy condition}",
    eprint = "1402.2694",
    archivePrefix = "arXiv",
    primaryClass = "gr-qc",
    doi = "10.1016/j.physletb.2014.05.026",
    journal = "Phys. Lett. B",
    volume = "734",
    pages = "44--48",
    year = "2014"
}

@article{Toshmatov:2014nya,
    author = "Toshmatov, Bobir and Ahmedov, Bobomurat and Abdujabbarov, Ahmadjon and Stuchlik, Zdenek",
    title = "{Rotating Regular Black Hole Solution}",
    eprint = "1404.6443",
    archivePrefix = "arXiv",
    primaryClass = "gr-qc",
    doi = "10.1103/PhysRevD.89.104017",
    journal = "Phys. Rev. D",
    volume = "89",
    number = "10",
    pages = "104017",
    year = "2014"
}

@article{Azreg-Ainou:2014pra,
    author = {Azreg-A{\"\i}nou, Mustapha},
    title = "{Generating rotating regular black hole solutions without complexification}",
    eprint = "1405.2569",
    archivePrefix = "arXiv",
    primaryClass = "gr-qc",
    doi = "10.1103/PhysRevD.90.064041",
    journal = "Phys. Rev. D",
    volume = "90",
    number = "6",
    pages = "064041",
    year = "2014"
}

@article{Azreg-Ainou:2014aqa,
    author = {Azreg-A{\"\i}nou, Mustapha},
    title = "{From static to rotating to conformal static solutions: Rotating imperfect fluid wormholes with(out) electric or magnetic field}",
    eprint = "1401.4292",
    archivePrefix = "arXiv",
    primaryClass = "gr-qc",
    doi = "10.1140/epjc/s10052-014-2865-8",
    journal = "Eur. Phys. J. C",
    volume = "74",
    number = "5",
    pages = "2865",
    year = "2014"
}

@article{Jusufi:2019caq,
    author = "Jusufi, Kimet and Jamil, Mubasher and Chakrabarty, Hrishikesh and Wu, Qiang and Bambi, Cosimo and Wang, Anzhong",
    title = "{Rotating regular black holes in conformal massive gravity}",
    eprint = "1911.07520",
    archivePrefix = "arXiv",
    primaryClass = "gr-qc",
    doi = "10.1103/PhysRevD.101.044035",
    journal = "Phys. Rev. D",
    volume = "101",
    number = "4",
    pages = "044035",
    year = "2020"
}

@article{Kumar:2020owy,
    author = "Kumar, Rahul and Ghosh, Sushant G.",
    title = "{Rotating black holes in $4D$ Einstein-Gauss-Bonnet gravity and its shadow}",
    eprint = "2003.08927",
    archivePrefix = "arXiv",
    primaryClass = "gr-qc",
    doi = "10.1088/1475-7516/2020/07/053",
    journal = "JCAP",
    volume = "07",
    pages = "053",
    year = "2020"
}

@article{Ghosh:2014pba,
    author = "Ghosh, Sushant G.",
    title = "{A nonsingular rotating black hole}",
    eprint = "1408.5668",
    archivePrefix = "arXiv",
    primaryClass = "gr-qc",
    doi = "10.1140/epjc/s10052-015-3740-y",
    journal = "Eur. Phys. J. C",
    volume = "75",
    number = "11",
    pages = "532",
    year = "2015"
}

@article{Kumar:2020hgm,
    author = "Kumar, Rahul and Ghosh, Sushant G. and Wang, Anzhong",
    title = "{Gravitational deflection of light and shadow cast by rotating Kalb-Ramond black holes}",
    eprint = "2001.00460",
    archivePrefix = "arXiv",
    primaryClass = "gr-qc",
    doi = "10.1103/PhysRevD.101.104001",
    journal = "Phys. Rev. D",
    volume = "101",
    number = "10",
    pages = "104001",
    year = "2020"
}

@article{Ghosh:2021clx,
    author = "Ghosh, Sushant G. and walia, Rahul Kumar",
    title = "{Rotating black holes in general relativity coupled to nonlinear electrodynamics}",
    eprint = "2109.13031",
    archivePrefix = "arXiv",
    primaryClass = "gr-qc",
    doi = "10.1016/j.aop.2021.168619",
    journal = "Annals Phys.",
    volume = "434",
    pages = "168619",
    year = "2021"
}

@article{Ali:2024ssf,
    author = "Ali, Heena and Islam, Shafqat Ul and Ghosh, Sushant G.",
    title = "{Shadows and parameter estimation of rotating quantum corrected black holes and constraints from EHT observation of M87* and Sgr A*}",
    eprint = "2410.09198",
    archivePrefix = "arXiv",
    primaryClass = "gr-qc",
    doi = "10.1016/j.jheap.2025.100367",
    journal = "JHEAp",
    volume = "47",
    pages = "100367",
    year = "2025"
}

@article{Kumar:2022vfg,
    author = "Kumar, Jitendra and Islam, Shafqat Ul and Ghosh, Sushant G.",
    title = "{Loop Quantum Gravity motivated multihorizon rotating black holes}",
    eprint = "2209.13562",
    archivePrefix = "arXiv",
    primaryClass = "gr-qc",
    doi = "10.1088/1475-7516/2022/11/032",
    journal = "JCAP",
    volume = "11",
    pages = "032",
    year = "2022"
}

@article{Kerr:1963ud,
    author = "Kerr, Roy P.",
    title = "{Gravitational field of a spinning mass as an example of algebraically special metrics}",
    doi = "10.1103/PhysRevLett.11.237",
    journal = "Phys. Rev. Lett.",
    volume = "11",
    pages = "237--238",
    year = "1963"
}

@article{Zhang:2022nnj,
    author = "Zhang, Jing and Xie, Yi",
    title = {{Gravitational lensing by a black-bounce-Reissner{\textendash}Nordstr{\"o}m spacetime}},
    doi = "10.1140/epjc/s10052-022-10441-7",
    journal = "Eur. Phys. J. C",
    volume = "82",
    number = "5",
    pages = "471",
    year = "2022"
}

@article{Dymnikova:2015hka,
    author = "Dymnikova, Irina and Galaktionov, Evgeny",
    title = "{Regular rotating electrically charged black holes and solitons in non-linear electrodynamics minimally coupled to gravity}",
    eprint = "1510.01353",
    archivePrefix = "arXiv",
    primaryClass = "gr-qc",
    doi = "10.1088/0264-9381/32/16/165015",
    journal = "Class. Quant. Grav.",
    volume = "32",
    number = "16",
    pages = "165015",
    year = "2015"
}

@article{Berej:2006cc,
    author = "Berej, Waldemar and Matyjasek, Jerzy and Tryniecki, Dariusz and Woronowicz, Mariusz",
    title = "{Regular black holes in quadratic gravity}",
    eprint = "hep-th/0606185",
    archivePrefix = "arXiv",
    doi = "10.1007/s10714-006-0270-9",
    journal = "Gen. Rel. Grav.",
    volume = "38",
    pages = "885--906",
    year = "2006"
}

@article{Junior:2015fya,
    author = "Junior, Ednaldo L. B. and Rodrigues, Manuel E. and Houndjo, Mahouton J. S.",
    title = "{Regular black holes in $f(T)$ Gravity through a nonlinear electrodynamics source}",
    eprint = "1503.07857",
    archivePrefix = "arXiv",
    primaryClass = "gr-qc",
    doi = "10.1088/1475-7516/2015/10/060",
    journal = "JCAP",
    volume = "10",
    pages = "060",
    year = "2015"
}

@article{Rodrigues:2019xrc,
    author = "Rodrigues, Manuel E. and de Sousa Silva, Marcos V.",
    title = "{Regular multihorizon black holes in $f(G)$ gravity with nonlinear electrodynamics}",
    eprint = "1906.06168",
    archivePrefix = "arXiv",
    primaryClass = "gr-qc",
    doi = "10.1103/PhysRevD.99.124010",
    journal = "Phys. Rev. D",
    volume = "99",
    number = "12",
    pages = "124010",
    year = "2019"
}

@article{deSousaSilva:2018kkt,
    author = "de Sousa Silva, Marcos V. and Rodrigues, Manuel E.",
    title = "{Regular black holes in $f(G)$ gravity}",
    eprint = "1808.05861",
    archivePrefix = "arXiv",
    primaryClass = "gr-qc",
    doi = "10.1140/epjc/s10052-018-6122-4",
    journal = "Eur. Phys. J. C",
    volume = "78",
    number = "8",
    pages = "638",
    year = "2018"
}

@article{Junior:2020zdt,
    author = "Junior, Ednaldo L. B. and Rodrigues, Manuel E. and de Sousa Silva, Marcos V.",
    title = "{Regular black holes in Rainbow Gravity}",
    eprint = "2002.04410",
    archivePrefix = "arXiv",
    primaryClass = "gr-qc",
    doi = "10.1016/j.nuclphysb.2020.115244",
    journal = "Nucl. Phys. B",
    volume = "961",
    pages = "115244",
    year = "2020"
}

@article{Cano:2020ezi,
    author = "Cano, Pablo A. and Murcia, {\'A}ngel",
    title = {{Resolution of Reissner-Nordstr{\"o}m singularities by higher-derivative corrections}},
    eprint = "2006.15149",
    archivePrefix = "arXiv",
    primaryClass = "hep-th",
    reportNumber = "IFT-UAM/CSIC-20-99",
    doi = "10.1088/1361-6382/abd923",
    journal = "Class. Quant. Grav.",
    volume = "38",
    number = "7",
    pages = "075014",
    year = "2021"
}

@book{Visser:1995cc,
    author    = "Visser, Matt",
    title     = "{Lorentzian Wormholes: From Einstein to Hawking}",
    publisher = "Amer Inst of Physics",
    year      = "1995",
    isbn      = "978-1-56396-653-8"
}

@book{Lobo:2017cay,
    editor = "Lobo, Francisco S. N.",
    title = "{Wormholes, Warp Drives and Energy Conditions}",
    eprint = "2103.05610",
    archivePrefix = "arXiv",
    primaryClass = "gr-qc",
    doi = "10.1007/978-3-319-55182-1",
    isbn = "978-3-319-55181-4, 978-3-319-85588-2, 978-3-319-55182-1",
    publisher = "Springer",
    series = "Fundamental Theories of Physics",
    volume = "189",
    year = "2017"
}

@article{Simpson:2019cer,
    author = "Simpson, Alex and Martin-Moruno, Prado and Visser, Matt",
    title = "{Vaidya spacetimes, black-bounces, and traversable wormholes}",
    eprint = "1902.04232",
    archivePrefix = "arXiv",
    primaryClass = "gr-qc",
    doi = "10.1088/1361-6382/ab28a5",
    journal = "Class. Quant. Grav.",
    volume = "36",
    number = "14",
    pages = "145007",
    year = "2019"
}

@article{Bronnikov:2021liv,
    author = "Bronnikov, Kirill A. and Konoplya, Roman A. and Pappas, Thomas D.",
    title = "{General parametrization of wormhole spacetimes and its application to shadows and quasinormal modes}",
    eprint = "2102.10679",
    archivePrefix = "arXiv",
    primaryClass = "gr-qc",
    doi = "10.1103/PhysRevD.103.124062",
    journal = "Phys. Rev. D",
    volume = "103",
    number = "12",
    pages = "124062",
    year = "2021"
}

@article{Guerrero:2021ues,
    author = "Guerrero, Merce and Olmo, Gonzalo J. and Rubiera-Garcia, Diego and G{\'o}mez, Diego S{\'a}ez-Chill{\'o}n",
    title = "{Shadows and optical appearance of black bounces illuminated by a thin accretion disk}",
    eprint = "2105.15073",
    archivePrefix = "arXiv",
    primaryClass = "gr-qc",
    doi = "10.1088/1475-7516/2021/08/036",
    journal = "JCAP",
    volume = "08",
    pages = "036",
    year = "2021"
}

@article{Lobo:2020ffi,
    author = "Lobo, Francisco S. N. and Rodrigues, Manuel E. and de Sousa Silva, Marcos V. and Simpson, Alex and Visser, Matt",
    title = "{Novel black-bounce spacetimes: wormholes, regularity, energy conditions, and causal structure}",
    eprint = "2009.12057",
    archivePrefix = "arXiv",
    primaryClass = "gr-qc",
    doi = "10.1103/PhysRevD.103.084052",
    journal = "Phys. Rev. D",
    volume = "103",
    number = "8",
    pages = "084052",
    year = "2021"
}

@article{Franzin:2021vnj,
    author = "Franzin, Edgardo and Liberati, Stefano and Mazza, Jacopo and Simpson, Alex and Visser, Matt",
    title = "{Charged black-bounce spacetimes}",
    eprint = "2104.11376",
    archivePrefix = "arXiv",
    primaryClass = "gr-qc",
    doi = "10.1088/1475-7516/2021/07/036",
    journal = "JCAP",
    volume = "07",
    pages = "036",
    year = "2021"
}

@article{Mazza:2021rgq,
    author = "Mazza, Jacopo and Franzin, Edgardo and Liberati, Stefano",
    title = "{A novel family of rotating black hole mimickers}",
    eprint = "2102.01105",
    archivePrefix = "arXiv",
    primaryClass = "gr-qc",
    doi = "10.1088/1475-7516/2021/04/082",
    journal = "JCAP",
    volume = "04",
    pages = "082",
    year = "2021"
}

@article{Bronnikov:2024izh,
    author = "Bronnikov, Kirill A.",
    title = "{Regular black holes as an alternative to black bounce}",
    eprint = "2404.14816",
    archivePrefix = "arXiv",
    primaryClass = "gr-qc",
    doi = "10.1103/PhysRevD.110.024021",
    journal = "Phys. Rev. D",
    volume = "110",
    number = "2",
    pages = "024021",
    year = "2024"
}

@article{Amir:2015pja,
    author = "Amir, Muhammed and Ghosh, Sushant G.",
    title = "{Rotating Hayward{\textquoteright}s regular black hole as particle accelerator}",
    eprint = "1503.08553",
    archivePrefix = "arXiv",
    primaryClass = "gr-qc",
    doi = "10.1007/JHEP07(2015)015",
    journal = "JHEP",
    volume = "07",
    pages = "015",
    year = "2015"
}

@article{Alencar:2024yvh,
    author = "Alencar, G. and Bronnikov, Kirill A. and Rodrigues, Manuel E. and S{\'a}ez-Chill{\'o}n G{\'o}mez, Diego and de S. Silva, Marcos V.",
    title = "{On black bounce space-times in non-linear electrodynamics}",
    eprint = "2403.12897",
    archivePrefix = "arXiv",
    primaryClass = "gr-qc",
    doi = "10.1140/epjc/s10052-024-13119-4",
    journal = "Eur. Phys. J. C",
    volume = "84",
    number = "7",
    pages = "745",
    year = "2024"
}

@article{Tzikas:2018cvs,
    author = "Tzikas, Athanasios G.",
    title = "{Bardeen black hole chemistry}",
    eprint = "1811.01104",
    archivePrefix = "arXiv",
    primaryClass = "gr-qc",
    doi = "10.1016/j.physletb.2018.11.036",
    journal = "Phys. Lett. B",
    volume = "788",
    pages = "219--224",
    year = "2019"
}

@article{Cheng:2021hoc,
    author = "Cheng, Xiao-Tong and Xie, Yi",
    title = "{Probing a black-bounce, traversable wormhole with weak deflection gravitational lensing}",
    doi = "10.1103/PhysRevD.103.064040",
    journal = "Phys. Rev. D",
    volume = "103",
    number = "6",
    pages = "064040",
    year = "2021"
}

@article{Cardoso:2014sna,
  author  = {Cardoso, V. and Crispino, L. C. B. and Macedo, C. F. B. and Okawa, H. and Pani, P.},
  title   = {Light rings as observational evidence for event horizons},
  journal = {Phys. Rev. D},
  volume  = {90},
  pages   = {044069},
  year    = {2014}
}

@article{Cunha:2017qtt,
  author  = {Cunha, P. V. P. and Berti, E. and Herdeiro, C. A. R.},
  title   = {Light-Ring Stability for Ultracompact Objects},
  journal = {Phys. Rev. Lett.},
  volume  = {119},
  pages   = {251102},
  year    = {2017}
}

@article{Bozza:2008ev,
    author = "Bozza, V.",
    title = "{A Comparison of approximate gravitational lens equations and a proposal for an improved new one}",
    eprint = "0807.3872",
    archivePrefix = "arXiv",
    primaryClass = "gr-qc",
    doi = "10.1103/PhysRevD.78.103005",
    journal = "Phys. Rev. D",
    volume = "78",
    pages = "103005",
    year = "2008"
}

@article{Nedkova:2013msa,
    author = "Nedkova, Petya G. and Tinchev, Vassil K. and Yazadjiev, Stoytcho S.",
    title = "{Shadow of a rotating traversable wormhole}",
    eprint = "1307.7647",
    archivePrefix = "arXiv",
    primaryClass = "gr-qc",
    doi = "10.1103/PhysRevD.88.124019",
    journal = "Phys. Rev. D",
    volume = "88",
    number = "12",
    pages = "124019",
    year = "2013"
}

@article{Darwin:1959zz,
    author = "Darwin, C.",
    title = "{The Gravity Field of a Particle}",
    doi = "https://doi.org/10.1098/rspa.1959.0015",
    journal = "Proceedings of the Royal Society of London Series A",
    volume = "249",
    pages = "180",
    year = "1959"
}

@article{Luminet:1979nyg,
    author = "Luminet, J. -P.",
    title = "{Image of a spherical black hole with thin accretion disk}",
    journal = "Astron. Astrophys.",
    volume = "75",
    pages = "228--235",
    year = "1979"
}

@article{Bozza:2003cp,
    author = "Bozza, V. and Mancini, L.",
    title = "{Time delay in black hole gravitational lensing as a distance estimator}",
    eprint = "gr-qc/0305007",
    archivePrefix = "arXiv",
    doi = "10.1023/B:GERG.0000010486.58026.4f",
    journal = "Gen. Rel. Grav.",
    volume = "36",
    pages = "435--450",
    year = "2004"
}

@article{Bozza:2002zj,
  author  = {Bozza, V.},
  title   = {Gravitational lensing in the strong field limit},
  journal = {Phys. Rev. D},
  volume  = {66},
  pages   = {103001},
  year    = {2002}
}

@article{Virbhadra:1999nm,
    author = "Virbhadra, K. S. and Ellis, George F. R.",
    title = "{Schwarzschild black hole lensing}",
    eprint = "astro-ph/9904193",
    archivePrefix = "arXiv",
    doi = "10.1103/PhysRevD.62.084003",
    journal = "Phys. Rev. D",
    volume = "62",
    pages = "084003",
    year = "2000"
}

@article{Teo:2003ltt,
    author = "Teo, Edward",
    title = "{Spherical Photon Orbits Around a Kerr Black Hole}",
    doi = "10.1023/A:1026286607562",
    journal = "Gen. Rel. Grav.",
    volume = "35",
    number = "11",
    pages = "1909--1926",
    year = "2003"
}

@article{Bardeen:1973tla,
    author = "Bardeen, J. M.",
    editor = "DeWitt, C{\'e}cile and DeWitt, Bryce Seligman",
    title = "{Timelike and null geodesics in the Kerr metric}",
    journal = "Proceedings, Ecole d'Et{\'e} de Physique Th{\'e}orique: Les Astres Occlus : Les Houches, France, August, 1972, 215-240",
    pages = "215--240",
    year = "1973"
}

@article{Shaikh:2018oul,
    author = "Shaikh, Rajibul and Banerjee, Pritam and Paul, Suvankar and Sarkar, Tapobrata",
    title = "{A novel gravitational lensing feature by wormholes}",
    eprint = "1811.08245",
    archivePrefix = "arXiv",
    primaryClass = "gr-qc",
    doi = "10.1016/j.physletb.2018.12.030",
    journal = "Phys. Lett. B",
    volume = "789",
    pages = "270--275",
    year = "2019",
    note = "[Erratum: Phys.Lett.B 791, 422--423 (2019)]"
}

@article{Godani:2021atr,
    author = "Godani, Nisha and Samanta, Gauranga C.",
    title = "{Gravitational lensing effect in traversable wormholes}",
    eprint = "2105.08517",
    archivePrefix = "arXiv",
    primaryClass = "gr-qc",
    doi = "10.1016/j.aop.2021.168460",
    journal = "Annals Phys.",
    volume = "429",
    pages = "168460",
    year = "2021"
}

@article{Lupsasca:2024xhq,
    author = "Lupsasca, Alexandru and C{\'a}rdenas-Avenda{\~n}o, Alejandro and Palumbo, Daniel C. M. and Johnson, Michael D. and Gralla, Samuel E. and Marrone, Daniel P. and Galison, Peter and Tiede, Paul and Keeble, Lennox",
    title = "{The Black Hole Explorer: photon ring science, detection, and shape measurement}",
    eprint = "2406.09498",
    archivePrefix = "arXiv",
    primaryClass = "gr-qc",
    doi = "10.1117/12.3019437",
    journal = "Proc. SPIE Int. Soc. Opt. Eng.",
    volume = "13092",
    pages = "130926Q",
    year = "2024"
}

@article{Nandi:2006ds,
    author = "Nandi, Kamal Kanti and Zhang, Yuan-Zhong and Zakharov, Alexander V.",
    title = "{Gravitational lensing by wormholes}",
    eprint = "gr-qc/0602062",
    archivePrefix = "arXiv",
    doi = "10.1103/PhysRevD.74.024020",
    journal = "Phys. Rev. D",
    volume = "74",
    pages = "024020",
    year = "2006"
}

@article{Tsukamoto:2012xs,
    author = "Tsukamoto, Naoki and Harada, Tomohiro and Yajima, Kohji",
    title = "{Can we distinguish between black holes and wormholes by their Einstein ring systems?}",
    eprint = "1207.0047",
    archivePrefix = "arXiv",
    primaryClass = "gr-qc",
    reportNumber = "RUP-12-5",
    doi = "10.1103/PhysRevD.86.104062",
    journal = "Phys. Rev. D",
    volume = "86",
    pages = "104062",
    year = "2012"
}

@article{Tsukamoto:2016zdu,
    author = "Tsukamoto, Naoki and Harada, Tomohiro",
    title = "{Light curves of light rays passing through a wormhole}",
    eprint = "1607.01120",
    archivePrefix = "arXiv",
    primaryClass = "gr-qc",
    reportNumber = "RUP-16-21",
    doi = "10.1103/PhysRevD.95.024030",
    journal = "Phys. Rev. D",
    volume = "95",
    number = "2",
    pages = "024030",
    year = "2017"
}

@article{Tsukamoto:2016qro,
    author = "Tsukamoto, Naoki",
    title = "{Strong deflection limit analysis and gravitational lensing of an Ellis wormhole}",
    eprint = "1607.07022",
    archivePrefix = "arXiv",
    primaryClass = "gr-qc",
    doi = "10.1103/PhysRevD.94.124001",
    journal = "Phys. Rev. D",
    volume = "94",
    number = "12",
    pages = "124001",
    year = "2016"
}

@article{Shaikh:2018kfv,
    author = "Shaikh, Rajibul",
    title = "{Shadows of rotating wormholes}",
    eprint = "1803.11422",
    archivePrefix = "arXiv",
    primaryClass = "gr-qc",
    doi = "10.1103/PhysRevD.98.024044",
    journal = "Phys. Rev. D",
    volume = "98",
    number = "2",
    pages = "024044",
    year = "2018"
}

@article{Bronnikov:2018nub,
    author = "Bronnikov, K. A. and Baleevskikh, K. A.",
    title = "{On gravitational lensing by symmetric and asymmetric wormholes}",
    eprint = "1812.05704",
    archivePrefix = "arXiv",
    primaryClass = "gr-qc",
    doi = "10.1134/S020228931901002X",
    journal = "Grav. Cosmol.",
    volume = "25",
    number = "1",
    pages = "44--49",
    year = "2019"
}

@article{Olmo:2023lil,
    author = "Olmo, Gonzalo J. and Rosa, Joao Luis and Rubiera-Garcia, Diego and Saez-Chillon Gomez, Diego",
    title = "{Shadows and photon rings of regular black holes and geonic horizonless compact objects}",
    eprint = "2302.12064",
    archivePrefix = "arXiv",
    primaryClass = "gr-qc",
    reportNumber = "IPARCOS-UCM-23-111",
    doi = "10.1088/1361-6382/aceacd",
    journal = "Class. Quant. Grav.",
    volume = "40",
    number = "17",
    pages = "174002",
    year = "2023"
}

@article{Solodukhin:2025opw,
    author = "Solodukhin, Sergey N. and Tagiev, Vagif",
    title = "{Exploring the landscape of black hole mimickers}",
    eprint = "2511.03879",
    archivePrefix = "arXiv",
    primaryClass = "gr-qc",
    doi = "10.1103/zc1z-sw9b",
    journal = "Phys. Rev. D",
    volume = "113",
    number = "6",
    pages = "064017",
    year = "2026"
}

@article{Zhang:2024sgs,
    author = "Zhang, Jing and Xie, Yi",
    title = "{Strong deflection gravitational lensing by the marginally unstable photon spheres of a wormhole}",
    doi = "10.1103/PhysRevD.109.043032",
    journal = "Phys. Rev. D",
    volume = "109",
    number = "4",
    pages = "043032",
    year = "2024"
}

@article{Virbhadra:1998dy,
    author = "Virbhadra, K. S. and Narasimha, D. and Chitre, S. M.",
    title = "{Role of the scalar field in gravitational lensing}",
    eprint = "astro-ph/9801174",
    archivePrefix = "arXiv",
    journal = "Astron. Astrophys.",
    volume = "337",
    pages = "1--8",
    year = "1998"
}

@article{Virbhadra:2002ju,
    author = "Virbhadra, K. S. and Ellis, G. F. R.",
    title = "{Gravitational lensing by naked singularities}",
    doi = "10.1103/PhysRevD.65.103004",
    journal = "Phys. Rev. D",
    volume = "65",
    pages = "103004",
    year = "2002"
}

@article{Shaikh:2018lcc,
    author = "Shaikh, Rajibul and Kocherlakota, Prashant and Narayan, Ramesh and Joshi, Pankaj S.",
    title = "{Shadows of spherically symmetric black holes and naked singularities}",
    eprint = "1802.08060",
    archivePrefix = "arXiv",
    primaryClass = "astro-ph.HE",
    doi = "10.1093/mnras/sty2624",
    journal = "Mon. Not. Roy. Astron. Soc.",
    volume = "482",
    number = "1",
    pages = "52--64",
    year = "2019"
}

@article{Johnson:2024ttr,
    author = "Johnson, Michael D. and others",
    title = "{The Black Hole Explorer: motivation and vision}",
    eprint = "2406.12917",
    archivePrefix = "arXiv",
    primaryClass = "astro-ph.IM",
    doi = "10.1117/12.3019835",
    journal = "Proc. SPIE Int. Soc. Opt. Eng.",
    volume = "13092",
    pages = "130922D",
    year = "2024"
}

@article{Shaikh:2019itn,
  author  = {Shaikh, R. and Banerjee, P. and Paul, S. and Sarkar, T.},
  title   = {Analytical approach to strong gravitational lensing from ultracompact objects},
  journal = {Phys. Rev. D},
  volume  = {99},
  pages   = {104040},
  year    = {2019}
}

@article{Tsukamoto:2016jzh,
  author  = {Tsukamoto, N.},
  title   = {Deflection angle in the strong deflection limit},
  journal = {Phys. Rev. D},
  volume  = {95},
  pages   = {064035},
  year    = {2017}
}

@article{Akiyama:2019eap,
  author  = {Akiyama, K. and others},
  collaboration = {Event Horizon Telescope},
  title   = {First M87 Event Horizon Telescope Results. VI. The Shadow and Mass of the Central Black Hole},
  journal = {Astrophys. J. Lett.},
  volume  = {875},
  pages   = {L6},
  year    = {2019}
}

@article{Simpson:2018tsi,
  author  = {Simpson, A. and Visser, M.},
  title   = {Black-bounce to traversable wormhole},
  journal = {JCAP},
  volume  = {02},
  pages   = {042},
  year    = {2019}
}

@article{Nascimento:2020ime,
  author  = {Nascimento, J. R. and Petrov, A. Y. and Porfirio, P. J. and Soares, A. R.},
  title   = {Gravitational lensing in a black-bounce traversable wormhole spacetime},
  journal = {Phys. Rev. D},
  volume  = {102},
  pages   = {044021},
  year    = {2020}
}

@article{Tsukamoto:2020bjm,
  author  = {Tsukamoto, N.},
  title   = {Gravitational lensing in the Simpson--Visser black-bounce spacetime},
  journal = {Phys. Rev. D},
  volume  = {103},
  pages   = {024033},
  year    = {2021}
}

@article{Ovgun:2020yuv,
    author = {{\"O}vg{\"u}n, Ali},
    title = "{Weak Deflection Angle of Black-bounce Traversable Wormholes Using Gauss-Bonnet Theorem in the Dark Matter Medium}",
    eprint = "2011.04423",
    archivePrefix = "arXiv",
    primaryClass = "gr-qc",
    doi = "10.20944/preprints202008.0512.v1",
    journal = "Turk. J. Phys.",
    volume = "44",
    number = "5",
    pages = "465--471",
    year = "2020"
}

@article{EventHorizonTelescope:2022wkp,
    author = "Akiyama, Kazunori and others",
    collaboration = "Event Horizon Telescope",
    title = "{First Sagittarius A* Event Horizon Telescope Results. I. The Shadow of the Supermassive Black Hole in the Center of the Milky Way}",
    eprint = "2311.08680",
    archivePrefix = "arXiv",
    primaryClass = "astro-ph.HE",
    doi = "10.3847/2041-8213/ac6674",
    journal = "Astrophys. J. Lett.",
    volume = "930",
    number = "2",
    pages = "L12",
    year = "2022"
}

@article{Ayzenberg:2023hfw,
    author = "Ayzenberg, D. and others",
    title = "{Fundamental physics opportunities with future ground-based mm/sub-mm VLBI arrays}",
    eprint = "2312.02130",
    archivePrefix = "arXiv",
    primaryClass = "astro-ph.HE",
    doi = "10.1007/s41114-025-00057-0",
    journal = "Living Rev. Rel.",
    volume = "28",
    number = "1",
    pages = "4",
    year = "2025",
    note = "[Erratum: Living Rev.Rel. 28, 7 (2025)]"
}

@article{Tsupko:2021yca,
    author = "Tsupko, Oleg Yu.",
    title = "{Deflection of light rays by a spherically symmetric black hole in a dispersive medium}",
    eprint = "2102.00553",
    archivePrefix = "arXiv",
    primaryClass = "gr-qc",
    doi = "10.1103/PhysRevD.103.104019",
    journal = "Phys. Rev. D",
    volume = "103",
    number = "10",
    pages = "104019",
    year = "2021"
}

@article{Ditta:2023aw,
title = {Particle dynamics and weak gravitational lensing around nonlinear electrodynamics black hole},
author = {Allah Ditta and Xia Tiecheng and Farruh Atamurotov and G. Mustafa and M.M. Aripov},
journal = {Chinese Journal of Physics},
volume = {83},
pages = {664-679},
year = {2023},
issn = {0577-9073}
}

@article{Ditta:2023rhr,
    author = "Ditta, Allah and Tiecheng, Xia and Atamurotov, Farruh and Hussain, Ibrar and Mustafa, G.",
    title = "{Particle dynamics, black hole shadow and weak gravitational lensing in the f (Q) theory of gravity}",
    doi = "10.1088/1572-9494/ad0e05",
    journal = "Commun. Theor. Phys.",
    volume = "75",
    number = "12",
    pages = "125404",
    year = "2023"
}

@article{Feng:2024jeq,
    author = "Feng, Yihu and Ditta, Allah and Mustafa, G. and Maurya, S. K. and Mahmood, Asif and Atamurotov, Farruh",
    title = "{Non-commutative Schwarzschild black hole surrounded by Perfect fluid dark matter: Plasma lensing and thermodynamics analysis}",
    doi = "10.1016/j.nuclphysb.2024.116713",
    journal = "Nucl. Phys. B",
    volume = "1008",
    pages = "116713",
    year = "2024"
}

@article{Ashraf:2025cnz,
    author = "Ashraf, Asifa and Ditta, Allah and Bouzenada, Abdelmalek and Maurya, S. K. and Abd-Elmonem, Assmaa and Suoliman, Nagat A. A. and Channuie, Phongpichit",
    title = "{Plasma lensing, epicyclic oscillations, particle collision, and thermal fluctuations around a short-hairy black hole}",
    doi = "10.1016/j.dark.2025.101836",
    journal = "Phys. Dark Univ.",
    volume = "48",
    pages = "101836",
    year = "2025"
}
\bibliographystyle{apsrev4-2}
\end{document}